\documentclass[journal]{IEEEtran}

\usepackage{times}
\usepackage{epsfig}
\usepackage{amssymb}
\usepackage{subfigure}
\usepackage{bbding}
\usepackage{xcolor}
\usepackage{threeparttable}
\usepackage{booktabs}
\usepackage{algorithm}
\usepackage{algorithmic}
\usepackage{url}

\newcommand{\eg}{\textit{e}.\textit{g}.}
\newcommand{\ie}{\textit{i}.\textit{e}.}

\usepackage{listings}
\usepackage{warpcol}
\usepackage{makecell}
\usepackage{amsthm,amsmath,amssymb}
\usepackage{mathrsfs}
\newcommand{\etal}{\textit{et al}.}

\usepackage{graphicx, color, multirow, amsmath}

\RequirePackage{latexsym,amsmath,amssymb}

\ifCLASSINFOpdf
\else
\fi

\begin{document}

\title{Subjective and Objective Quality Assessment of Non-Uniformly Distorted Omnidirectional Images}

\author{Jiebin Yan,
    Jiale Rao,
    Xuelin Liu,
    Yuming Fang,~\IEEEmembership{Senior~Member,~IEEE},
    Yifan Zuo,~\IEEEmembership{Member,~IEEE},
    Weide Liu

\thanks{This work was supported in part by the National Key Research and Development Program of China under Grant 2023YFE0210700, in part by the National Natural Science Foundation of China under Grants 62132006, 62461028, 62441203, and 62311530101, in part by the Natural Science Foundation of Jiangxi Province of China under Grants 20223AEI91002, 20232BAB202001, 20224ACB212005, and 20242BAB26014, and in part by the project funded by China Postdoctoral Science Foundation under Grant 2024T170364. (Corresponding author: Yuming Fang).}

\thanks{J. Yan, J. Rao, X. Liu, Y. Fang, and Y. Zuo are with the School of Computing and Artificial Intelligence, Jiangxi University of Finance and Economics, Nanchang 330032, China (e-mail: \{jiebinyan, jialerao, xuelinliu-bill\}@foxmail.com, fa0001ng@e.ntu.edu.sg, kenny0410@126.com).}
\thanks{W. Liu is with Harvard Medical School, Harvard University, USA (e-mail: weide001@e.ntu.edu.sg).}}


\maketitle

\begin{abstract}
Omnidirectional image quality assessment (OIQA) has been one of the hot topics in IQA with the continuous development of VR techniques, and achieved much success in the past few years. However, most studies devote themselves to the uniform distortion issue, \ie, all regions of an omnidirectional image are perturbed by the ``same amount'' of noise, while ignoring the non-uniform distortion issue, \ie, partial regions undergo ``different amount'' of perturbation with the other regions in the same omnidirectional image. Additionally, nearly all OIQA models are verified on the platforms containing a limited number of samples, which largely increases the over-fitting risk and therefore impedes the development of OIQA. To alleviate these issues, we elaborately explore this topic from both subjective and objective perspectives. Specifically, we construct a large OIQA database containing 10,320 non-uniformly distorted omnidirectional images, each of which is generated by considering quality impairments on one or two camera len(s). Then we meticulously conduct psychophysical experiments and delve into the influence of both holistic and individual factors (\ie, distortion range and viewing condition) on omnidirectional image quality. Furthermore, we propose a perception-guided OIQA model for non-uniform distortion by adaptively simulating users' viewing behavior. Experimental results demonstrate that the proposed model outperforms state-of-the-art methods. The source code is available at \url{https://github.com/RJL2000/OIQAND}.
\end{abstract}


\IEEEpeerreviewmaketitle

\section{Introduction}
\label{sec:intr}

\IEEEPARstart{W}{ith} the emergence of the metaverse concept, the processing of omnidirectional images (OIs) have gained significant attention from both academia and industry community. OI, also referred to as \(360^\circ\) image or panoramic image, is a crucial medium in Virtual Reality (VR) applications, and has the capability to capture all-encompassing natural scenes from the real world, and users view them in a three-dimensional fashion through head-mounted displays (HMDs). Users can immerse themselves in a freely explorative experience, gaining a sense of being present in the depicted environment. However, the introduced distortion would easily degrade the visual quality of OIs and affect users' quality of experience (QoE)~\cite{IQASurvey}, and it is of great significance in assessing perceptual quality of OIs accurately due to the applications on algorithm improvement and multimedia system optimization~\cite{ding2021comparison,fang2021superpixel}. Omnidirectional image quality assessment (OIQA) includes subjective OIQA and objective OIQA, where the former generally refers to conduct psychophysical experiments on the quality of OIs and investigate the influence of various quality-aware factors, such as the internal factors in producing stage and external factors in playback stage~\cite{yan2022subjective}. The outputs of psychophysical experiments, \ie, subjective databases, always act as benchmarks for testing objective OIQA models and guiding their further improvement. Objective OIQA models aim to automatically predict the perceptual quality of OIs. The first type of OIQA methods~\cite{s-psnr,ws-psnr,cpp-psnr,ws-ssim, gao2020quality} mainly rely on mature 2D-IQA methods, such as PSNR and SSIM~\cite{ssim}, however, their performance is far from being satisfactory since they don't consider the spherical characteristic of OIs and users' viewing behavior. The second type of OIQA methods firstly project OIs to different formats, such as equirectangular projection (ERP), cubemap (CMP), and pyramid projections (PYM)~\cite{Jiang2021tip,li2023mfan,zhou2023perception}, and then extract information from one of these formats (or combine these formats together) to represent the quality of OIs. However, these methods also ignore users' viewing behavior, and thus their performance is sub-optimal. The third type of models, \ie, viewport-based~\cite{zhou2022no}, ~\cite{tian2022vsoiqe}, ~\cite{tian2023viewport}, can solve the above problems well by modeling users' viewing behaviors~\cite{Sui,fang2022perceptual} or empirically extracting viewports (\ie, field of view) according to the prior knowledge, \ie, users are prone to look at the regions nearly the equator.

\begin{figure}[t]
\centering
\includegraphics[width=0.9\linewidth]{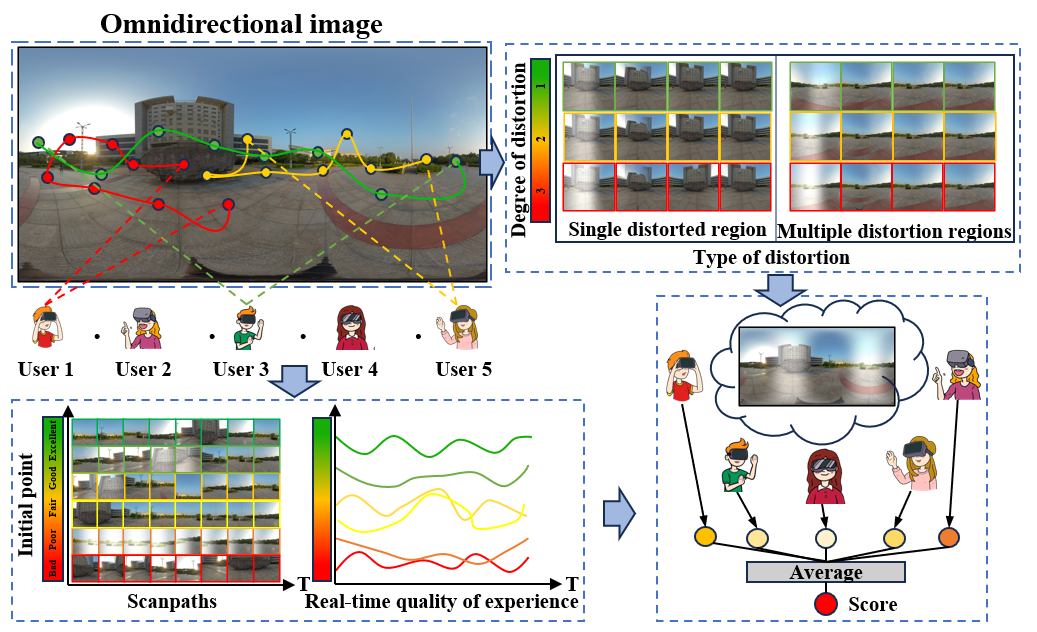}
\caption{A visual example of how users explore omnidirectional images.}
\label{fig:intro}
\end{figure} 

Although OIQA has gained progressive success in terms of subjective databases and objective models, however, there are still at least two weaknesses. First, the number of image samples (most OIQA databases only contain hundreds of images) is relatively limited compared with those recently released 2D-IQA databases~\cite{fang2020perceptual,ma2016waterloo} and OI databases for other task~\cite{wu2022view}, which contain tens of thousands of images. As the key \textit{fuel} for training deep models, the large number of image samples plays irreplaceable role in promoting the development of IQA models~\cite{fang2020perceptual,ma2016waterloo}. Limited number of OIs is obviously an obstacle in the way of development, since it brings the high-risk over-fitting problem which makes us hard to draw the conclusion on which model/module performs better than the others. Second, almost all of existing OIQA databases except the JUFE~\cite{fang2022perceptual} database concentrate on globally uniform distortion, \ie, all regions in an OI are perturbed by the same type of distortion with same level, while ignoring locally non-uniform distortion. Actually, those in-wild OIs are easily disturbed by a variety of non-uniform distortions. From the imaging perspective (\ie, generally, an omnidirectional camera is composed of multiple lenses, \eg, an Insta Pro2 camera has six lenses), non-uniform distortion would be introduced in many cases, \eg, fast moving objects lead to motion blur in one or two lenses, and uneven exposure may result in inconsistent depth of field. From the generation perspective (\ie, an OI is generated from several 2D fisheye images by a stitching algorithm), the generation algorithms would introduce discontinuity between two fisheye images, which brings local distorted regions in the generated OIs~\cite{zhou2023quality}. As shown in Fig.~\ref{fig:intro}, users tend to exhibit distinct navigation scanpaths for omnidirectional images due to users' different viewing conditions (\ie, users browse omnidirectional images from different starting points) and images' various distortion situations (\ie, those images are subjected to varying degrees of disturbance), especially when those images suffer from non-uniform distortion, leading to difficulty in modeling the perceived quality of omnidirectional images.

To solve the above problems, in this paper, we delve into the non-uniform distortion issue in OIQA from both subjective and objective perspectives. Specifically, we construct a large-scale OIQA database with rich content and complicated distortion, where we simulate real situations by artificially disturbing one or two fisheye images, leading to locally distributed distortion in OIs. Besides, to be consistent with users' actual viewing behaviors, we investigate the influence of viewing condition on the QoE of OIs by randomly setting viewing starting points, instead of setting fixed starting points~\cite{fang2022perceptual}. Then, we propose a perception-guided OIQA model for non-uniform distortion named OIQAND, where we elaborately design two parallel modules, \ie, a multi-scale feature fusion module and a distortion adaptive perception module, for better capturing distortion-related features, and an additional module, \ie, the viewport-wise perceptual quality prediction module, for simulating the recency effect of the human visual system (HVS)~\cite{Sui,recency}. The experimental results show that the proposed model outperforms existing state-of-the-art models.

\begin{table*}[h]\normalsize%
\caption{Summary of the existing OIQA databases and the proposed JUFE-10K database. ERP represents the equirectangular projection. GB represents Gaussian blur. GN represents Gaussian noise. BD represents brightness discontinuities. ST represents stitching distortion. ``No. Ref/Dist'' column stands for the number of reference and distorted OIs, respectively. ``Resolution'' column is the largest resolution in the database.}

\begin{center}

\scalebox{0.78}{
\begin{tabular}{p{3.4cm}<{\centering}|m{1cm}<{\centering} m{2cm}<{\centering} m{1.5cm}<{\centering} m{3cm}<{\centering}  m{2.5cm}<{\centering} m{6.5cm}<{\centering}}
\toprule
Database & Year & No.Ref/Dist & Range & Resolution & Format/Projection & Distortion type\\[2pt]
\midrule
OIQA~\cite{OIQA2018}  & 2018 & 16/320 &  [1, 10]  & 13,320$\times$6,660 &  PNG, JPG/ERP & Uniform: (JPEG, JP2K, GB, GN)\\[5pt]
CVIQ~\cite{CVIQ} & 2018 & 16/528 & [0, 100]  & 4,096$\times$2,048   & PNG/ERP & Uniform: (JPEG, H.264/AVC, H.265/HEVC) \\[5pt]
MVAQD~\cite{Jiang2021tip} & 2021 & 15/300 & [1, 5] & 5,780$\times$2,890  & BMP/ERP   & Uniform: (JPEG, JP2K, HEVC, WN, GB) \\[5pt]
IQA-ODI~\cite{IQA-ODI} & 2021 & 120/960 & [0, 100]  & 7,680$\times$3,840  & JPG /ERP  & Uniform: (JPEG, Projection) \\[5pt]
SOLID~\cite{xu2018SOLID}& 2018 & 6/276 & [1, 5] & 8,192$\times$4,096 & PNG/ERP & Uniform: (JPEG, BPG)\\[5pt]
LIVE 3D VR IQA~\cite{LIVE3D}& 2019 & 15/450& [0, 100] & 4,096$\times$2,048 &PNG/ERP &Uniform: (GB, GN, Downsampling, VP9 compression, H.265); Non-uniform: ST \\[5pt]
ISIQA~\cite{madhusudana2019subjective} & 2019 & 26/264 & [0, 100] & 9,270$\times$1,680   & PNG/ERP & Non-uniform: Stitching\\[5pt]
CROSS~\cite{li2019cross}   & 2019 & 292/2,044  & [0, 100] & 5,792$\times$2,894   & PNG/ERP & Non-uniform: Stitching\\[5pt]
Sui~\etal~\cite{Sui}  & 2021 & 36/36  & [1, 5] & 7,680$\times$3,840              & PNG/ERP & Uniform: H.265; Non-uniform: ST\\[5pt]
JUFE~\cite{fang2022perceptual}   & 2022 & 258/1,032  & [1, 5] & 8,192$\times$4,096  & PNG/ERP & Non-uniform: (GB, GN, BD, ST)\\[5pt]
\midrule
JUFE-10K   & 2024 & 430/10,320  & [1, 5] & 8,192$\times$4,096   & PNG/ERP & Non-uniform: (GB, GN, BD, ST)\\[4pt]
\bottomrule
\end{tabular}}
\end{center}

\label{tab_existingDB}
\end{table*}

In summary, our contributions are three folds:
\begin{itemize}

\item A large-scale OIQA database with locally distributed non-uniform distortion is established, consisting of 10,320 OIs generated by simulating the realistic applications. To the best of knowledge, the proposed database is the largest database to date for non-uniform distortion, offering a new test bed for OIQA models.   

\item An extensive psychophysical experiment is conducted, where we collect both subjective quality ratings and viewing behaviors (\ie, eye and head movement, which are abbreviated by EM and HM, respectively) without explicitly setting viewing conditions, and we analyze the relationship among viewing conditions, distortion situations and subjects' QoE detailedly.

\item A novel perception-guided OIQA model that considers users' viewing behavior and the recency effect of the HVS is proposed. The experimental results show that the proposed model achieves better performance than the compared models.

\end{itemize}

\section{Related Work}
\label{sec:related}

In this section, we first introduce subjective OIQA databases and then review objective 2D-IQA and OIQA models.

\subsection{Subjective OIQA Databases}

Duan~\etal~\cite{OIQA2018} constructed an OIQA database, which contains four types of distortion: JPEG compression, JPEG2000 compression, Gaussian noise, and Gaussian blur. The image resolution ranges from 11,332×5,666 to 13,320×6,660. The single-stimulus (SS) method is adopted to collect subjective scores in the form of mean opinion score (MOS). In addition, they recorded subjects' HM and EM data. In order to explore the influence of popular compression coding techniques on the quality of OIs, Sun~\etal~\cite{CVIQ} constructed an OIQA database containing JPEG, H.264/AVC, and H.265/HEVC compression distortions, and the SS method is adopted in the subjective experiment. Jiang~\etal~\cite{Jiang2021tip} constructed an OIQA database containing five distortion types and four distortion levels, where image resolution is 5,780×2,890 and quality scores range from 1 to 5. To study the effect of projection methods, Yang~\etal~\cite{IQA-ODI} constructed an OIQA database, which contains JPEG compression distortion and the distortion introduced by projection. Xu~\etal~\cite{xu2018SOLID} established a stereoscopic OIQA database and collected quality score for each distorted OI. It is worth nothing that a stereoscopic OI consists of a left- and a right-view. Chen~\etal~\cite{LIVE3D} constructed another stereoscopic OIQA database with 450 distorted OIs, which are generated by adding six distortion types with five distortion levels to 15 reference images. Madhusudana~\etal~\cite{madhusudana2019subjective} and Li~\etal~\cite{li2019cross} constructed two OIQA databases containing only stitching distortions, whose number of distorted OIs are 264 and 2,044, respectively. In order to study the influence of viewing conditions on perceptual quality, Sui~\etal~\cite{Sui} conducted a case study by artificially setting two viewing initial points and two exploring times in the subjective experiment. Later, Fang~\etal~\cite{fang2022perceptual} constructed the first, to the best of our knowledge, OIQA database with the focus on non-uniform distortion, where two initial viewing points (good and bad) and two exploring times (5s and 15s) are pre-defined.  

On the whole, most of the databases described above focus on uniform distortion, such as compression distortion, or only stitching distortion. A few databases consider user behavior and non-uniform distortion. Moreover, the scale of the above databases is generally smaller than that of 2D-IQA databases and more likely to lead to the overfitting problem for model training. To alleviate these problems, we construct a new large-scale OIQA database (JUFE-10K) with non-uniform distortion. Table I summarizes the newly proposed JUFE-10K database and existing OIQA databases.

\begin{figure*}[]
\centering
\includegraphics[width=0.9\linewidth]{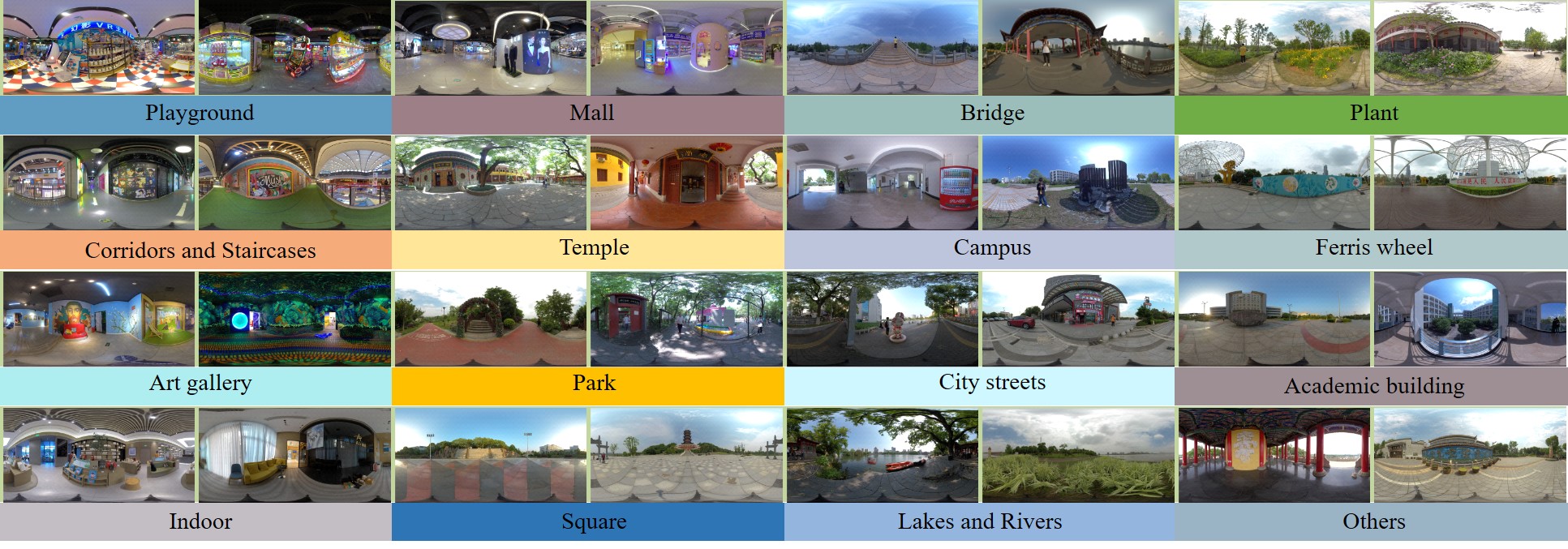}
\caption{Thumbnails of reference OIs in the proposed JUFE-10K database.}
\label{fig:database}
\end{figure*}

\subsection{Objective 2D-IQA Models}
\label{subsec:2d-iqa}

\begin{figure}[t]
\centering
\subfigure[OIQA]{
    \includegraphics[width=0.27\linewidth]{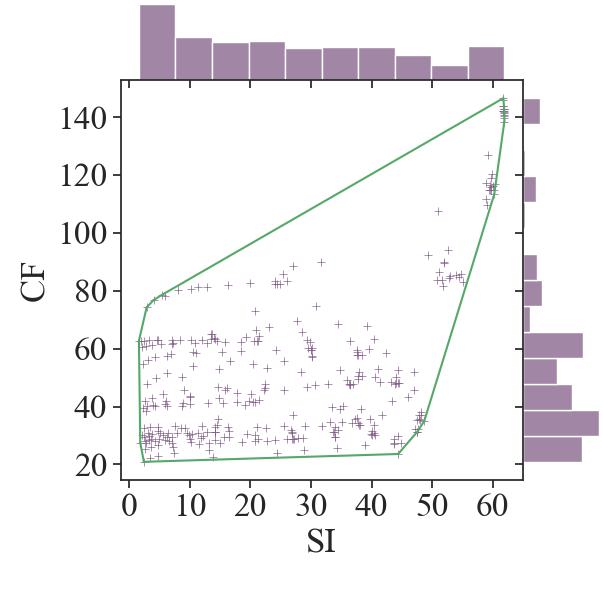}
}
\subfigure[CVIQ]{
    \includegraphics[width=0.27\linewidth]{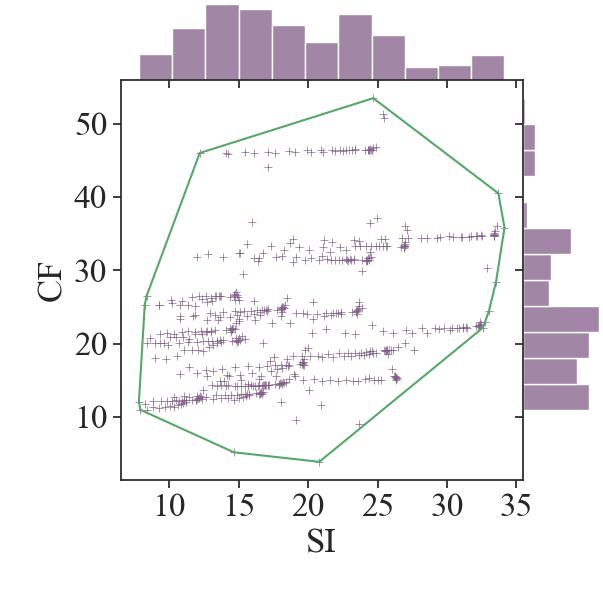}
}
\subfigure[JUFE]{
    \includegraphics[width=0.27\linewidth]{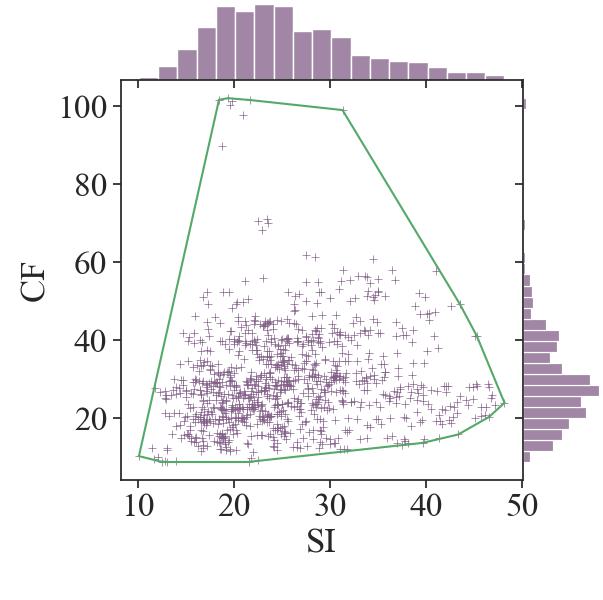}
}

\subfigure[LIVE 3D VR IQA]{
    \includegraphics[width=0.27\linewidth]{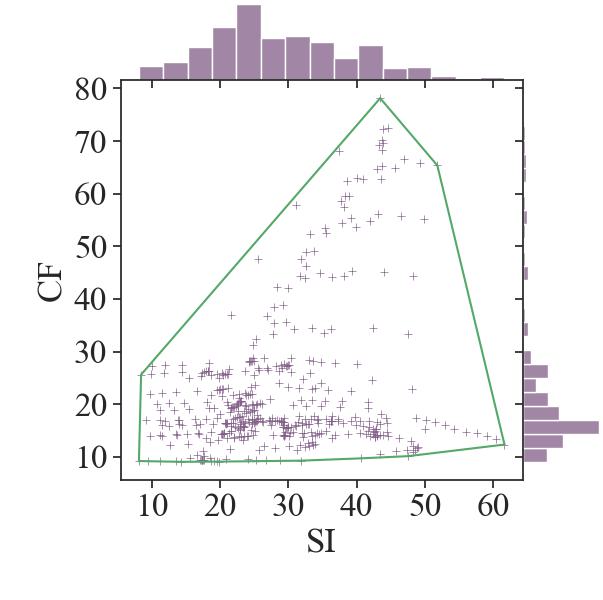}
}
\subfigure[SOLID]{
    \includegraphics[width=0.27\linewidth]{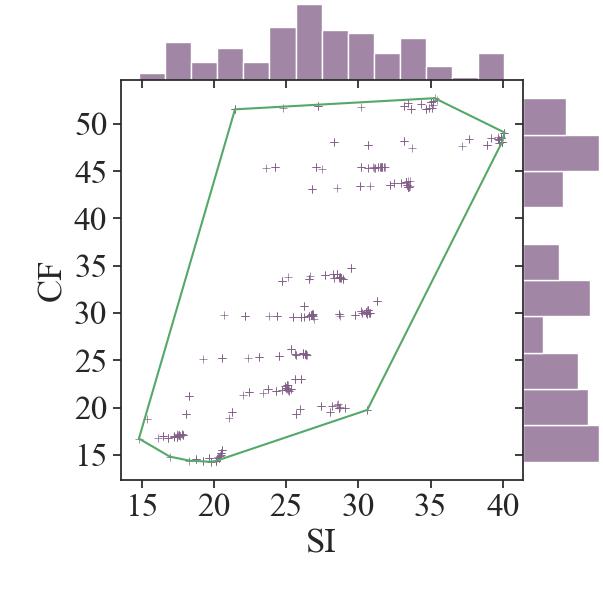}
}
\subfigure[JUFE-10K]{
    \includegraphics[width=0.27\linewidth]{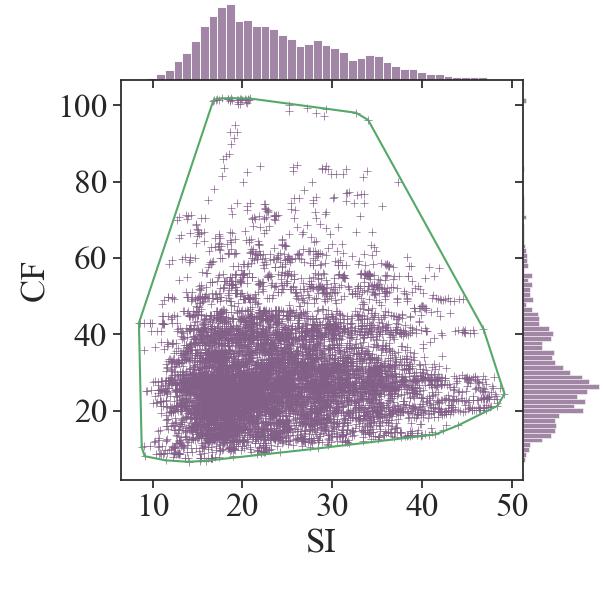}
}

\caption{The scatter diagram of spatial information (SI) and colorfulness (CF) of six OIQA databases.}
\label{fig:si_cf_index} 
\end{figure}

\begin{figure*}
\centering
  \includegraphics[width=1\linewidth]{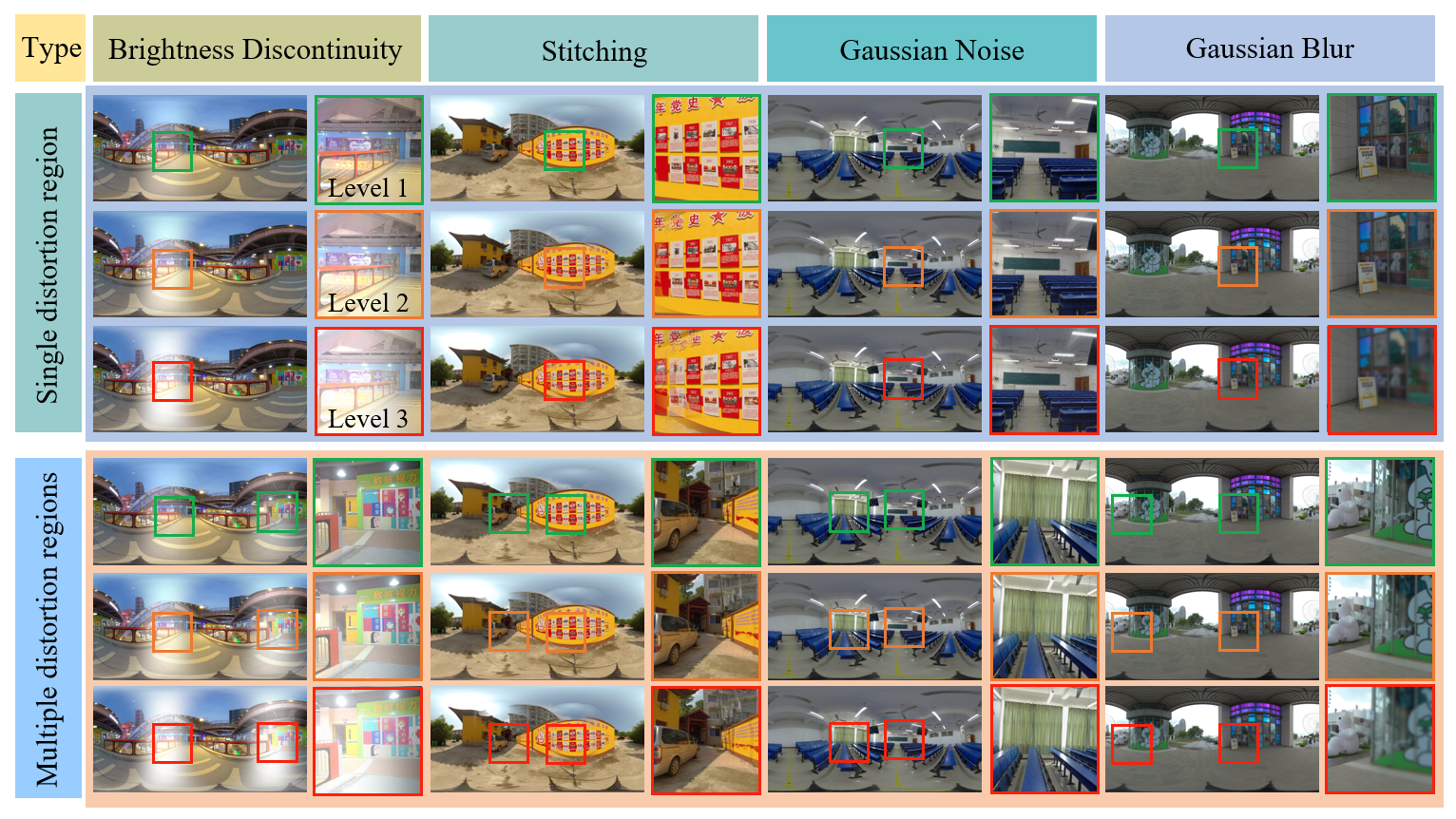}
\caption{Visual examples of non-uniform distortion. Zoom-in for better view.}
\label{fig:distype} 
\end{figure*}


According to the accessibility to reference information, IQA models can be divided into full-reference IQA (FR-IQA), reduced-reference IQA (RR-IQA), and no-reference IQA (NR-IQA) models. The FR-IQA and RR-IQA models require full and partial reference information when being used to predict image quality, and the NR-IQA models are more widely used in practice since they do not require reference information. These NR-IQA models can be further divided into traditional models and deep learning based models. The traditional NR-IQA method~\cite{fang2020blind} mainly consists of two stages: feature extraction and quality regression, and how to effectively extract features is the core of these methods. Mittal~\etal~\cite{mittal2012making} proposed a natural image quality evaluator (NIQE) model based on natural scene statistics (NSS), where the generalized Gaussian model is used to compute feature distribution difference between high-quality images and distorted images, and the difference is used to represent image quality. Fang~\etal~\cite{fang2017no} proposed to use histogram to represent the statistical quality-aware features of screen content images, rather than fitting feature distribution. Yan~\etal~\cite{yan2020no} introduced histogram to capture global quality change of the 3D synthesized images. 


Different from traditional NR-IQA methods, deep learning based NR-IQA methods learn the quality prediction model by automatic data-driven end-to-end optimization. Kang~\etal~\cite{kang2014convolutional} conducted an early study by introducing convolutional neural network (CNN) to IQA, where image patches are extracted as the input and the corresponding global quality scores are regarded as their ground-truth subjective ratings. Latter, many works focus on designing specific and efficient networks~\cite{yan2020blind}. Su~\etal~\cite{su2020blindly} proposed a self-adaptive hyper network for IQA by considering semantic features, whose main idea is that humans always evaluate image quality after recognizing image content. Madhusudana~\etal~\cite{madhusudana2022image} proposed to use contrastive learning to train a quality-aware network from scratch, and then fine-tune on the well-trained parameters to obtain the final IQA model. Yang~\etal~\cite{yang2022maniqa} proposed a multi-dimensional attention network for IQA, which consists of a basic network for extracting global and local features, a transposed attention block and a scale swin transformer block for enhancing global and local interactions, and a dual branch structure for quality prediction. Pan~\etal~\cite{pan2022vcrnet} introduced a restoration task to optimize the IQA model, where multi-level restoration features are used as the input of quality prediction module. Considering the ambiguity of subjective ratings, Zhang~\etal~\cite{zhang2021uncertainty} proposed a unified learning-to-rank framework by computing the probability that an image has higher quality than another image in the same pair. Zhang~\etal~\cite{zhang2023blind} designed a novel IQA model via vision-language correspondence, where the similarity between the embeddings of the pre-defined texture template and input image is used for multi-task learning. Yi~\etal~\cite{yi2023towards} proposed a network for artistic image aesthetic assessment with the consideration of generic aesthetic features and style-specific aesthetic features. 
 
\subsection{Objective OIQA Models}

Compared with 2D plane images, OIs have different storage formats, and different degrees of deformation would be introduced when being projected to 2D format. Therefore, 2D-IQA methods perform poorly on predicting the quality of OIs. In order to solve this problem, some researchers try to modify 2D-IQA algorithm to suit for OIs. Yu~\etal~\cite{s-psnr} proposed S-PSNR, which selects uniformly sampled points on the sphere to calculate PSNR. Sun \etal~\cite{ws-psnr} proposed WS-PSNR, which assigns weights to the pixels at different positions according to the relationship between spherical space and projection space. Zakharchenko~\etal~\cite{cpp-psnr} proposed CPP-PSNR by projecting OI to the Craster's parabolic projection (CPP) format. Analogously, S-SSIM~\cite{s-ssim} and WS-SSIM~\cite{ws-ssim} are designed based on SSIM~\cite{ssim}. As a mainstream solution of computer vision, deep learning has also been widely used in OIQA. According to models' input formats, those deep learning based OIQA models can be divided into two types: projection-based and viewport based. For the first type of OIQA models, the input OI is firstly converted to a projection space, such as ERP and CPP, and then fed into deep networks. Jiang~\etal~\cite{Jiang2021tip} proposed a perception driven OIQA method based on the CMP format, which enables omnidirectional viewing of on OI by projecting it into six interrelated face images, and designed three different schemes to calculate quality score. Zheng~\etal~\cite{zheng2020segmented} proposed to convert OI into the segmented spherical projection (SSP) format, and divide it into two polar regions and equatorial regions, and a variety of features are extracted from them. These features are input into the random forest regressor to obtain the quality score. Li~\etal~\cite{li2023mfan} proposed a multi-projection fusion attention network and designed an attention module to predict the scores of different projection spaces, then uses polynomial regression to obtain the quality scores. Zhou~\etal~\cite{zhou2023perception} designed a U-shaped transformer network combined with user perception, using multiple CMP image patches with different view directions as input, using the saliency map generation model to generate the visual features of the HVS and the U-former model to extract features, and jointly using the weighted average technique to predict the perceptual quality. 


The second type of OIQA models mainly simulate users' real viewing process, and take the viewport sequence as input. In this way, the visual content of the data is consistent with what the user actually sees. Sun~\etal~\cite{MC360IQA} proposed a multi-channel network for OIQA, where six parallel networks are used to extract the features of the viewport image, and the quality score is obtained after fusing multiple features. Xu~\etal~\cite{VGCN} proposed a viewport based graph convolutional neural network architecture to solve the interaction problem between different viewport images. Both viewport images and OIs are used to predict the final quality scores. Tofighi~\etal~\cite{jabbari2023st360iq} extracted viewport images by detecting salient sampling points, and input the extracted viewport images into the model to predict the perceptual quality of OIs. Fang~\etal~\cite{fang2022perceptual} combined different viewing conditions to study the impact of non-uniform distortion on user perceptual quality and designed an OIQA model based on CNN, which included a multi-scale feature extraction module based on CNN and a perceptual quality prediction module. In addition, viewing conditions were incorporated into the model. In order to model the user browsing process, Wu~\etal~\cite{wu2023assessor360} designed a recursive probability sampling method to extract viewport image sequences from OIs and input them into the network. 


\section{Subjective Experiment Methodology}

\subsection{Database Construction}
\label{subsec:da_con}

We extend the JUFE~\cite{fang2022perceptual} database from both content diversity and degradation complexity. As for content diversity, we firstly collect more than 600 images using an Insta360 Pro2 camera, and then carefully screen out 172 high-quality images by small-scale subjective judgement, forming the final high-quality image set with 430 OIs (note that 258 OIs are from the JUFE database). As shown in Fig.~\ref{fig:database}, these OIs can be classified into 16 categories, including playground, mall, bridge, plant,~\emph{etc}. In order to give an intuitive comparison with the existing OIQA databases, we show the distributions of Spatial Information (SI)~\cite{si_index} against Colorfulness (CF)~\cite{cf_index} in Fig.~\ref{fig:si_cf_index}, where we can clearly observe that the proposed JUFE-10K database has a noticeable increase in diversity compared to existing databases.

As for degradation complexity, we mainly consider the fact that OIs are easily affected by non-uniform distortion, such as camera sensors, object motion, uneven illumination intensity, and stitching algorithms, resulting in local quality discontinuity. To this end, we add distortion to one or two randomly selected nonadjacent fisheye image(s) and then use the Nuke\footnote{https://www.foundry.com/products/nuke-family/nuke} software to stitch all fisheye images to obtain the corresponding distorted OI. Same to the JUFE database, we consider four common types of non-uniform distortion: Gaussian noise (GN), Gaussian blur (GB), brightness discontinuities (BD), and stitching distortion (ST), each with 3 levels and 2 random ranges. Specifically, GN is used to simulate the sensor noise of InstaPro2 camera or noise introduced by the environment. GB is used to simulate the blur caused by motion and focal length in the real scenes. BD is used to simulate the brightness difference caused by uneven illumination intensity between multiple shots. ST is used to simulate the stitching distortion that occurs when the stitching algorithm cannot successfully deal with the inconsistency between multiple fisheye images in the later stitching process. We use the Nuke software to obtain the distorted OIs with the stitching distortion by adjusting the distortion parameters of one or two lenses. Distortion parameters are set to 0.5, 0.75, and 1. Finally, we obtain 10,320 non-uniformly distorted OIs. Visual examples of non-uniform distortion are shown in Fig.~\ref{fig:distype}.

\subsection{Psychophysical Experiment Design}
\label{subsec:sub_eva}
\subsubsection{Viewing Conditions}

Considering the randomness of the users' viewing process in real applications, we believe that artificially setting different viewing conditions is inconsistent with the actual viewing process. Therefore, instead of pre-defined viewing conditions, we randomly set viewing starting points and exploration times to simulate users' real viewing behavior. For each scene, a random viewing starting point means that each user may start viewing the OI from a low-quality region, a high-quality region, or a fair-quality region. To enable users to fully explore the whole OI, we set the time of a viewing process to 15 seconds.


\begin{figure}[]
\centering
\subfigure[Distribution and statistics of MOSs in term of distortion level]{
    \includegraphics[width=1\linewidth]{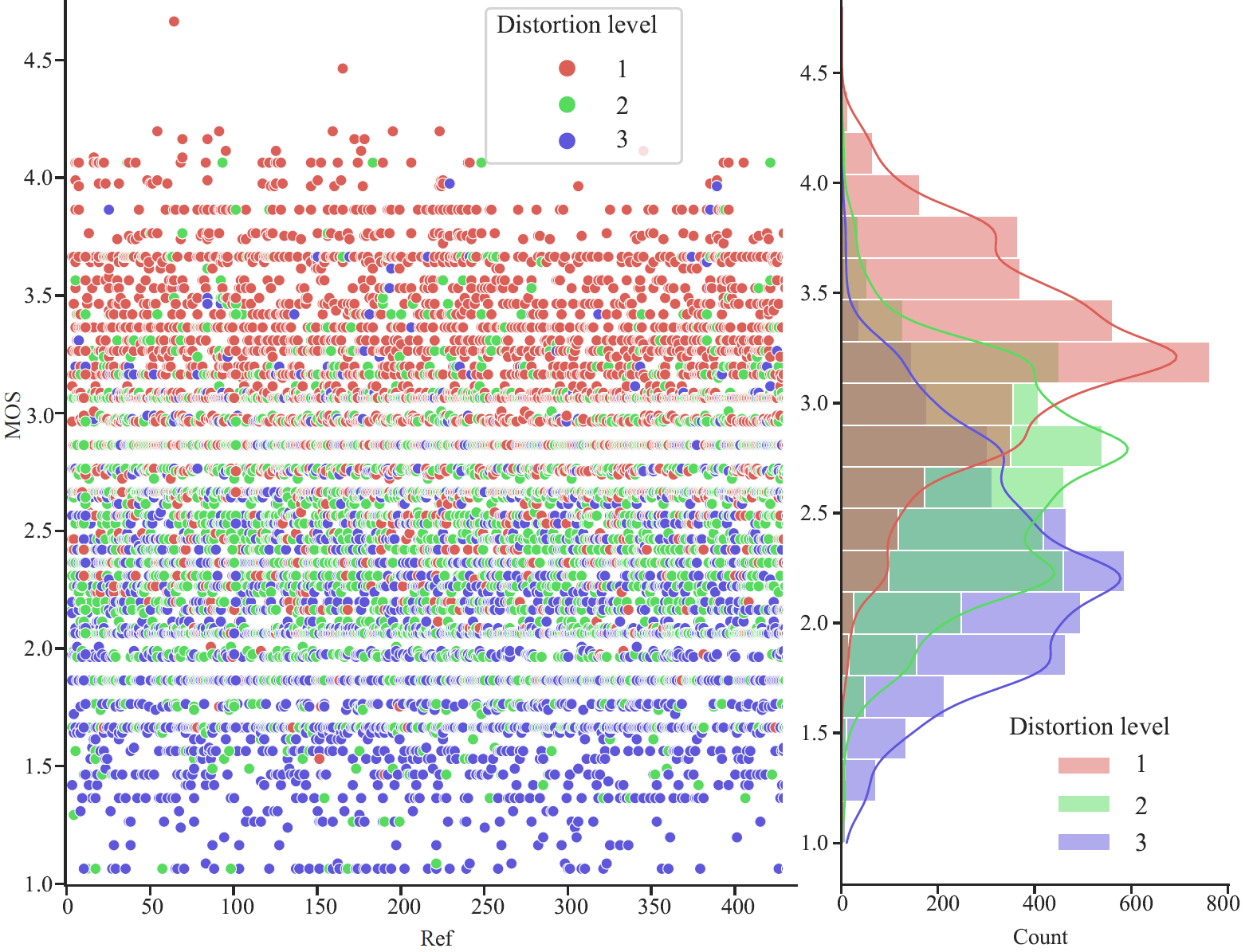}
    }\\
\subfigure[Distribution and statistics of MOSs in terms of distortion type]{
    \includegraphics[width=0.8\linewidth]{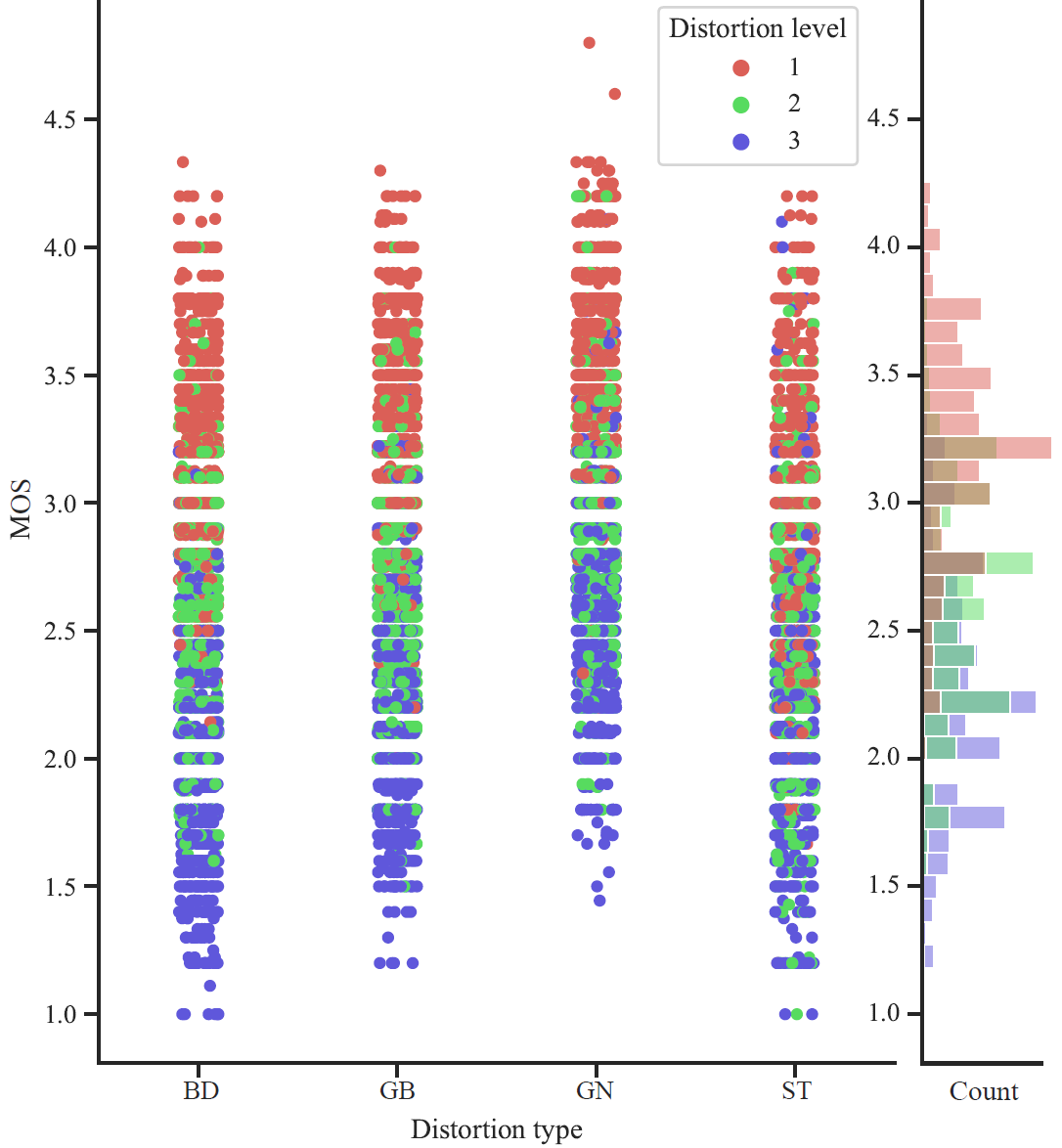}
    }\\
\subfigure[Boxplots of MOSs in terms of viewing initial point and distortion range]{
    \includegraphics[width=1\linewidth]{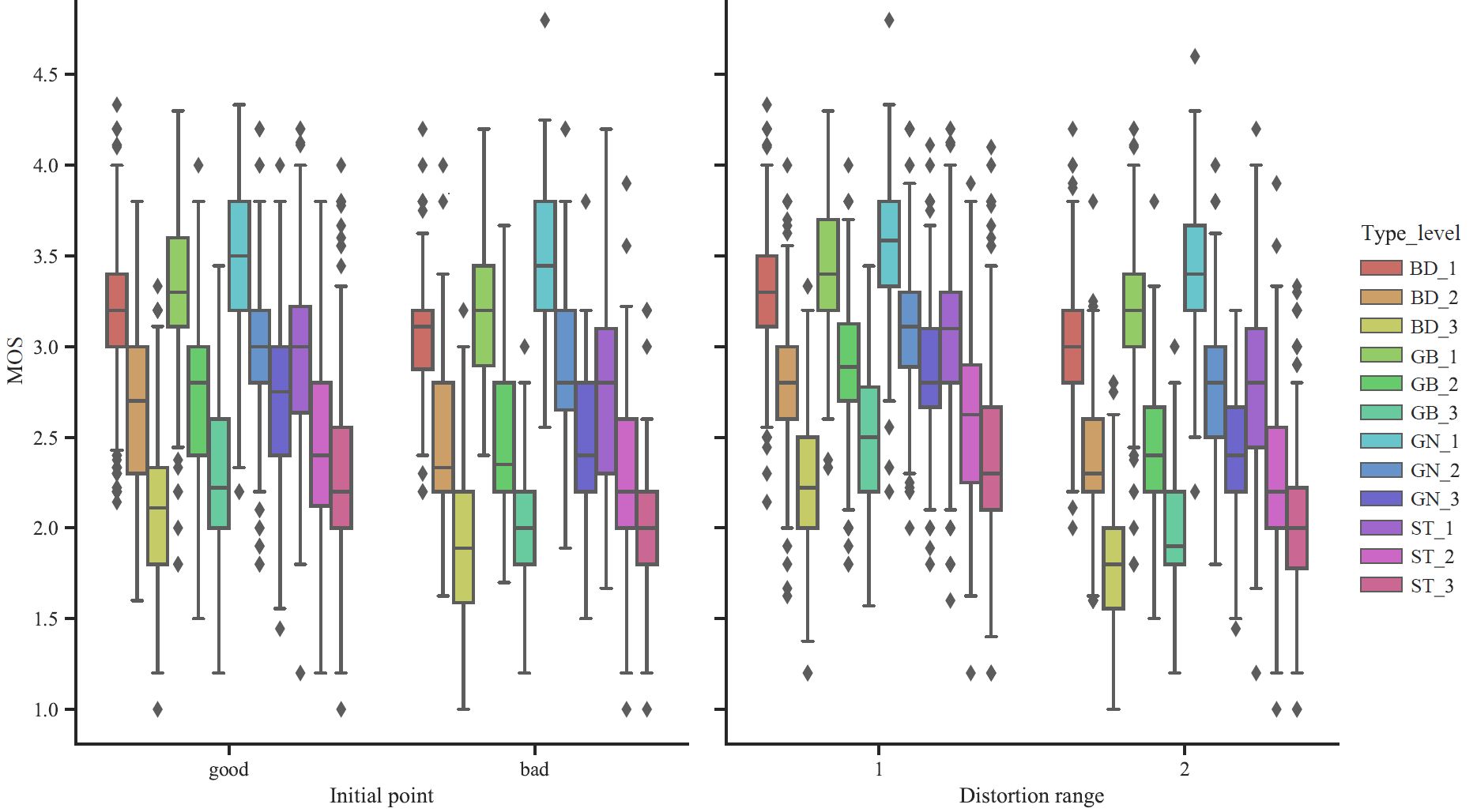}
    }
\caption{The statistic distributions of the quality of OIs from different perspectives.}
\label{fig:MOS_index} 
\end{figure}

\subsubsection{Subjective Methodology} 

We conduct subjective experiments using the SS method as recommended in the ITU-R BT.500.13~\cite{bt2002methodology}. Subjects are required to rate the quality of OIs in [1, 5]. The subjective experiment is conducted in a room without interference. The OIs are displayed in random order using the Unity software, and the image resolution is uniformly set to 8,192×4,096. HTC VIVE Pro Eye is used as the HMD, and the EM and HM data of the subjects are recorded by the built-in Tobii Pro eye tracking system. The experiment is conducted on a high-performance server equipped with AMD Ryzen 9 3950X 16-core CPU, 128 GB RAM, and GeForce RTX 2080 Ti GPU. A total of 172 subjects are invited to participate in the subjective experiment. In order to familiarize the subjects with the experimental process and exclude the subjects who are not suitable for the subjective experiment, we set a training set in advance to familiarize them with the distortion types and rating procedure. Image quality is set to 5 levels (``bad", ``poor", ``fair", ``good", and ``excellent"), which corresponds to 1-5 points. 

The whole subjective experiment is partitioned into two stages~\cite{yan2022subjective}. In the first stage, we divide 10,320 images into 80 groups, and each group is scored by five different subjects. When all 80 sets of images are scored, we calculate the variance of the scores for each OI and sort them from smallest to largest. We consider it credible that the score of images in the first quarter of variance is relatively stable, and we save this part of the score. In the second stage, we continue to group the images of the last three-quarters of variance into 61 groups for scoring again, and another 5 subjects are asked to rate these images. If any subject is tired or uncomfortable during the subjective experiment, they can stop to rest immediately. Totally, this subjective experiment costs more than 8 months.

\subsection{Subjective Data Processing and Analysis}
\label{subsec:ana_sub_data}
\subsubsection{Data Processing}
We use the method~\cite{bt2002methodology} to detect and remove abnormal subjects to further ensure the credibility of subjective data. Specifically, we perform $\beta_2$ test and check whether the distribution of scores for each image in each group is normal by calculating the mean $\mu$ and standard dedication $\sigma$ of scores. A parameter $n$ equals 2 (if normal) or $\sqrt{20}$ (if not normal), which is determined by the distribution of scores. We then calculate two statistics ($P_i$ and $Q_i$) for each image based on whether the score of image is within the confidence interval $[\mu - n\sigma, \mu +n\sigma]$. Next, we determine whether $\frac{P_{i}+Q_{i}}{N_{img}} $ is greater than 0.05 and $\left |\frac{P_{i}-Q_{i}}{P_{i}+Q_{i}}\right|$ is less than 0.3 ($N_{img}$ is the number of images). If both conditions are true, the subjective scores of this subject will be removed. In the end, we eliminate the quality scores rated by 37 abnormal subjects. Then, we calculate MOS of each image as follows:
\begin{eqnarray}
\centering 
MOS_{i}=\frac1{N}{\sum_{j=1}^{N}s_{ij}},
\label{eq:mos}
\end{eqnarray}
where $s_{ij}$ indicates the subjective score of image $i$ given by subject $j$; $N$ represents the number of subjects. 

\subsubsection{Data Analysis}

In Fig.~\ref{fig:MOS_index}, we visualize the statistic distributions of the quality of OIs from different perspectives. The left sub-figure of Fig.~\ref{fig:MOS_index} (a) shows the MOS distribution of all images. Obviously, we can observe that image scores decrease significantly with the increase of the distortion level. Meanwhile, the right sub-figure of Fig.~\ref{fig:MOS_index} (a) shows the quality statistics of OIs according to distortion level, and it can be seen that the statistics are close to the normal distribution. In order to test the influence of different distortion types on the perceived quality of OIs, Fig.~\ref{fig:MOS_index} (b) shows that different types of distortion present different MOS distributions at different distortion levels, and Gaussian noise has less influence on the perceived quality than the other types. The MOS coverage of brightness discontinuity, Gaussian blur, and stitching distortion are relatively close. From Fig.~\ref{fig:MOS_index} (c), we have several interesting observations. Firstly, from the left sub-figure, we can observe that the MOS of the good starting point is a little higher than that of the bad starting point, but the difference is not much. The reason may be that the subjects are allowed to have enough time to explore the whole OI, thus reducing the impact of different starting points. Secondly, we can see from the right figure that the MOS distribution of single distortion and multiple distortion is significantly different, and the MOS of multiple distortion is lower than that of single distortion, which is in line with the quality perception process of human beings.

\begin{figure}
\centering
  \includegraphics[width=1\linewidth]{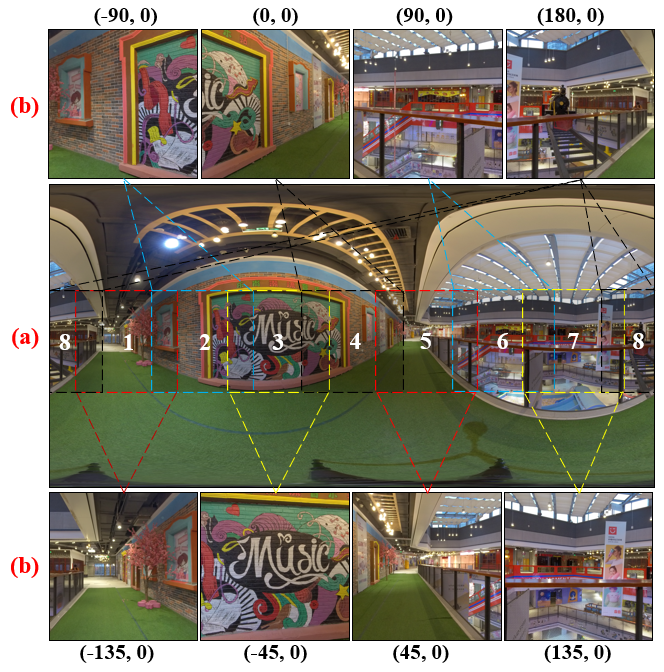}
\caption{Illustration of a source OI and its corresponding corresponding latitude and longitude viewport images. (a) Source image. (b) Viewport images.}
\label{fig:single_viewport} 
\end{figure}

\begin{figure*}
\centering
  \includegraphics[width=0.90\linewidth]{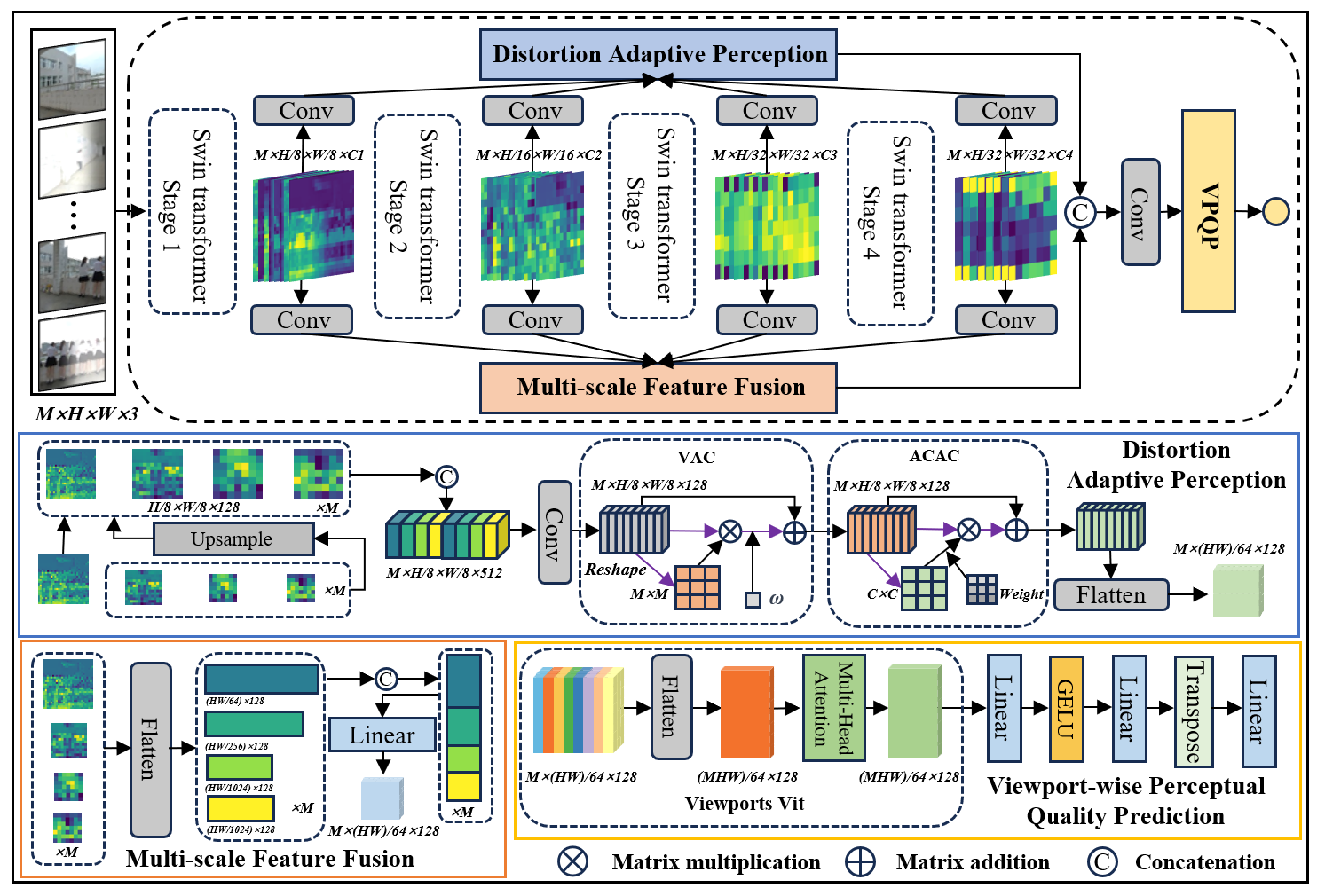}
\caption{The framework of the proposed method. It consists of three main modules, where the distortion perception module is used to adaptively adjust the importance of non-uniform distortion between different viewports, the multi-scale feature extraction module is used to extract multi-scale content features from multiple viewports, and the viewport-wise perceptual quality prediction module regresses the non-uniform distortion features of multiple viewports to quality score.}
\label{fig:framework} 
\end{figure*}

\section{The Proposed OIQA Model}
\label{sec:abff}

In this section, we propose a perception-guided OIQA model for Non-uniform Distortion namely OIQAND, and its architecture is shown in Fig.~\ref{fig:framework}. The design concept of OIQAND is to simulate users' viewing behavior, \emph{i.e.}, the content being watched can be regarded as a sequence of viewport (with about 90$^{\circ}$ field of view), and the overall quality of OIs will be affected by viewport-wise quality and their collective influence. Specifically, several viewport images are extracted as the input of this model, and a transformer-based backbone is utilized for multi-scale feature extraction. Then, a multi-scale feature fusion module and a distortion adaptive perception module are designed for capturing distortion in these viewport images, and a viewport-wise perceptual quality prediction module is used to map the deeply fused features to overall quality score.

\subsection{Viewport Extraction}
\label{subsec:sfem}
Generally, subjects explore an OI in a continuous process, and most subjects give priority to viewing the scene content near the equatorial region. Based on this priori knowledge, we believe that sampling viewports uniformly along the equatorial region is consistent with the actual process of continuous exploration. Therefore, we use rectilinear projection to generate viewport images, and $M$ viewport images (the default value of $M$ is set to 8) are extracted by uniform sampling from the equatorial region of each OI with 45-degree intervals. We will verify the effectiveness of this simple viewport generation in Section~\ref{subsec:as}. The specific procedure can be found in~\cite{JVETG1003}, and the extraction results are shown in Fig.~\ref{fig:single_viewport}. Given $N$ OIs, we can obtain a set of viewport images $VS=\left\{I_i,\left\{V_j,(\phi_j)\right\}_{j=1}^M\right\}_{i=1}^N$, where $VS$ denotes the viewport set, $I_i$ represents the $i$-th OI, $V_j$ is the $j$-th viewport image of the $i$-th image, $\phi_j$ represents the coordinate (\ie, latitude and longitude) of the $j$-th viewport.

\subsection{Model Construction}
\subsubsection{Multi-scale Feature Extraction}

As a new paradigm of deep learning, the transformer has strong versatility~\cite{vaswani2017attention}, and has been widely used in computer vision community. In order to efficiently capture the representation of each viewport from low-level features (edges, color variations, and textures) to high-level features (overall structure, semantics, and context), we use the swin transformer~\cite{liu2021swin} as our backbone $B$ to extract multi-scale features of viewports. As shown in Fig.~\ref{fig:framework}, feature maps $\mathbf{F}^1 \in \mathbb{R}^{ M \times \frac{H}{8} \times \frac{W}{8} \times C1 } $, $ \mathbf{F}^2 \in \mathbb{R}^{ M \times \frac{H}{16} \times \frac{W}{16} \times C2 } $, $ \mathbf{F}^3 \in \mathbb{R}^{ M \times \frac{H}{32} \times \frac{W}{32} \times C3 }$, and $ \mathbf{F}^4 \in \mathbb{R}^{ M \times \frac{H}{32} \times \frac{W}{32} \times C4} $ of the four stages can be obtained by
\begin{eqnarray}
\centering 
\mathbf{F}^s = \left\{(B(VS;\Theta_{B}))\right\}_{s=1}^4,
\label{eq:feature}
\end{eqnarray}
where $\Theta_{B}$ is the parameters of the backbone; $s$ denotes the index of four stages. To better fuse multi-scale features and avoid the dimension mismatch problem, we then use multiple 1×1 convolutions $\mathrm{W}_\Theta^1$, $\mathrm{W}_\Phi^2$, $\mathrm{W}_\Lambda^3$, and $\mathrm{W}_\Gamma^4$ to unify the channel numbers, and the resultant feature maps $\Theta^1 \in \mathbb{R}^{
 M \times \frac{H}{8} \times \frac{W}{8} \times \hat{C}}$, $\Phi^2 \in \mathbb{R}^{ M \times \frac{H}{16} \times \frac{W}{16} \times \hat{C}}$, $\Lambda^3 \in \mathbb{R}^{ M \times \frac{H}{32} \times \frac{W}{32} \times \hat{C}}$, and $\Gamma^4 \in \mathbb{R}^{ M \times \frac{H}{32} \times \frac{W}{32} \times \hat{C}}$ can be obtained by 
\begin{eqnarray}
\begin{split}
\centering 
&\Theta^1=\mathrm{W}_\Theta^1(\mathbf{F}^1),\Phi^2=\mathrm{W}_\Phi^2(\mathbf{F}^2),\\
&\Lambda^3=\mathrm{W}_\Lambda^3(\mathbf{F}^3),\Gamma^4=\mathrm{W}_\Gamma^4(\mathbf{F}^4),
\end{split}
\label{eq:feature}
\end{eqnarray}
where $\hat{C}$ = 128 represents the number of channels after unification; $M$ denotes the number of viewports.

\subsubsection{Viewport Distortion Perception}
\paragraph{Multi-scale Feature Fusion}
To effectively integrate the spatial structure information of feature maps from four different stages and better capture scene structure and patterns of viewports, we design a multi-scale feature fusion (MFF) module. Specifically, we flatten $\Theta^1$, $\Phi^2$, $\Lambda^3$, $\Gamma^4$ in $(H,\ W)$ dimension, and obtain the corresponding features $\Bar{\Theta} \in \mathbb{R}^{ M \times \frac{HW}{64} \times \hat{C}}$, $\Bar{\Phi} \in \mathbb{R}^{ M \times \frac{HW}{256} \times \hat{C}}$, $\Bar{\Lambda} \in \mathbb{R}^{ M \times \frac{HW}{1024} \times \hat{C}}$, and $\Bar{\Gamma} \in \mathbb{R}^{ M \times \frac{HW}{1024} \times \hat{C}}$. During the flattening process, the adjacent pixels of different feature maps still maintain the adjacent relationship and retain the relative position information. Then, $\Bar{\Theta}$, $\Bar{\Phi}$, $\Bar{\Lambda} $, and $\Bar{\Gamma} $ are concatenated in the flattened dimension to obtain the fused features $\mathbf{X}_{mf} \in \mathbb{R}^{ M \times N \times \hat{C}}$, and then they are sent into the linear transformation to obtain the final fused multi-scale features $ \mathbf{X}_{mff} \in \mathbb{R}^{ M \times \frac{HW}{64} \times \hat{C}}$. The computational process of the MFF module can be formulated as
\begin{eqnarray}
\centering 
\mathbf{X}_{mff}=W_1(\mathbf{X}_{mf})+b_1, \ \mathbf{X}_{mf}= \mathrm{Cat}(\Bar{\Theta}, \Bar{\Phi}, \Bar{\Lambda}, \Bar{\Gamma}),
\label{eq:feature}
\end{eqnarray}
where $W_1$ and $ b_1 $ are the parameters of the linear transformation.

\paragraph{Distortion Adaptive Perception}
Inspired by the feature pyramid network~\cite{lin2017feature, deng2022omnidirectional}, we design a distortion adaptive perception (DAP) module, we upsample these lower-resolution feature maps $\Phi^2$, $\Lambda^3 $, and $\Gamma^4 $ to match the size of $\Theta^1$, and concatenate them in the channel dimension to obtain the perceptual features $\mathbf{X}_\Psi \in \mathbb{R}^{ M \times \frac{H}{8} \times \frac{W}{8} \times C2}$. This process can introduce high-level semantic information into low-level features to achieve multi-scale information fusion and guide the model to understand viewport content. A 1×1 convolution $\mathrm{W}_\Psi$ is then used to reduce the number of dimensional channels and computational cost, and we can obtain the guidance feature map $\Psi \in \mathbb{R}^{ M \times \frac{H}{8} \times \frac{W}{8} \times \hat{C}}$. The whole process can be described as
\begin{eqnarray}
\centering 
\Psi=\mathrm{W}_\Psi(\mathbf{X}_\Psi),\ \mathbf{X}_\Psi= \mathrm{Cat}(\Theta^1,\mathrm{US}(\Phi^2,\Lambda^3,\Gamma^4)),
\label{eq:feature}
\end{eqnarray}
where $\mathrm{US}(\cdot)$ represents the upsampling operation. Considering the different proportions of non-uniform distortion in different viewports, we perform a series of viewport attention calculation (VAC) operations for non-uniform distortion by feat of the self-attention mechanism, which allows the model to dynamically allocate \textit{attentions} to different viewports. Specifically, we transform $\Psi$ into $\mathbf{\psi}_q \in \mathbb{R}^{M \times1}$ and $\mathbf{\psi}_k \in \mathbb{R}^{1\times M}$ by using global average pooling (GAP), fully connected layer and transpose operation, and multiply them and perform Softmax normalization to obtain viewport attention matrix $\mathbf{VAM} \in \mathbb{R}^{M \times M}$, and then reshape $\Psi$ into $\Psi_v \in \mathbb{R}^{M \times \frac{HW\hat{C}}{64}}$. The weighted viewport features $ \mathbf{V}_\Psi \in \mathbb{R}^{M \times \frac{HW\hat{C}}{64}}$ are obtained by matrix multiplication of $\mathbf{VAM}$ and $\Psi_v$. Finally, the feature map $\hat{\mathbf{V}} \in \mathbb{R}^{ M \times \frac{H}{8} \times \frac{W}{8} \times \hat{C}}$ can be obtained by reshaping the $\mathbf{V}_{\Psi}$ and adding it to $\Psi$ after scaling with a learnable factor $\omega$. The process can be described as
\begin{eqnarray}
\centering
\mathbf{VAM}^{ij} = \frac{\mathrm{exp}({\psi_q^i} \cdot {\psi_k^j})}{\sum_{m=1}^M \mathrm{exp}({\psi_q^i} \cdot {\psi_k^m})},
\label{eq:feature}
\end{eqnarray}

\begin{eqnarray}
\centering
\hat{\mathbf{V}} = \mathrm{R}(\sum_{m=1}^M ({\mathbf{VAM}^{im}} \cdot {\Psi_v^{mj}})) * \omega + \Psi,
\label{eq:feature}
\end{eqnarray}
where $*$ denotes scalar multiplication; $\psi_q^i$ denotes the $i$-th row element of $\psi_q$; $\psi_k^j$ and $\psi_k^m$ denote the $j$-th column element of $\psi_k$ and the $m$-th column element of $\psi_k$, respectively; $\Psi_v^{mj}$ denotes the $m$-th row and $j$-th column element of $\Psi_v$; $\mathrm{R}$ represents the reshape operation.

Similar to the VAC operation, we design an adaptive channel attention calculation (ACAC) module for guiding the model to dynamically learn the importance of different channel dimensions of each viewport and achieve adaptive fusion between channels. Specifically, $\hat{\mathbf{V}}$ is transformed into $\hat{\mathbf{V}}_q \in \mathbb{R}^{M\times \hat{C} }$ and $\hat{\mathbf{V}}_k \in \mathbb{R}^{ \hat{C}\times M }$ through the GAP and transposition operations. They are multiplied and normalized by the Softmax function to obtain channel attention matrix $\mathbf{CAM} \in \mathbb{R}^{\hat{C} \times \hat{C}}$. We introduce the channel adaptive weights $\mathbf{W} \in \mathbb{R}^{\hat{C} \times \hat{C}}$ multiplied with $\mathbf{CAM}$ and reshape $\hat{\mathbf{V}}$ into $\hat{\mathbf{V}}_v \in \mathbb{R}^{\hat{C} \times \frac{MHW}{64}}$.  The weighted channel features $ \dot{\mathbf{C}} \in \mathbb{R}^{\hat{C}\times \frac{MHW}{64}}$ are obtained by multiplying $\mathbf{CAM}$ and $\hat{\mathbf{V}}_v$. Finally, the $\dot{\mathbf{C}}$ is reshaped into $\mathbf{F}_{\mathbf{CAM}} \in \mathbb{R}^{ M \times \frac{H}{8} \times \frac{W}{8} \times \hat{C}} $ and added to $\hat{\mathbf{V}}$. The process can be formulated as

\begin{eqnarray}
\centering
\mathbf{CAM}^{ij} = \frac{\mathrm{exp}(\sum_{m=1}^M{\hat{\mathbf{V}}_k^{im}} \cdot {\hat{\mathbf{V}}_v^{mj}})}{\sum_{c=1}^C \mathrm{exp}(\sum_{m=1}^M{\hat{\mathbf{V}}_k^{im}} \cdot {\hat{\mathbf{V}}_v^{mc}})},
\label{eq:feature}
\end{eqnarray}

\begin{eqnarray}
\centering
\mathbf{F}_{\mathbf{CAM}} = \mathrm{R}(\sum_{c=1}^C ({ (\mathbf{CAM}\odot \mathbf{W}) ^{ic}}  \cdot {\hat{\mathbf{V}}_v^{cj}})) + \hat{\mathbf{V}},
\label{eq:feature}
\end{eqnarray}
where $\odot$ denotes element-wise multiplication; $\hat{\mathbf{V}}_k^{im}$ denotes the $i$-th row and $m$-th column element of $\hat{\mathbf{V}}_k$; $\hat{\mathbf{V}}_v^{mj}$ and $\hat{\mathbf{V}}_v^{mc}$ denote the $m$-th row and $j$-th column element of $\hat{\mathbf{V}}_v$ and the $m$-th row and $c$-th column element of $\hat{\mathbf{V}}_v$, respectively; $\hat{\mathbf{V}}_v^{cj}$ denotes the $c$-th row and $j$-th column element of $\hat{\mathbf{V}}_v$.

In order to improve the model's ability to capture and represent the diverse features of viewports, we fuse $\mathbf{X}_{mff}$ and $\mathbf{X}_{dap} \in \mathbb{R}^{ M \times \frac{HW}{64} \times \hat{C}}$, and get $ \Upsilon \in \mathbb{R}^{ M \times \frac{HW}{64} \times \hat{C}}$. Specifically, we concatenate the $\mathbf{X}_{dap}$ by flattening $\mathbf{F}_{\mathbf{CAM}}$ in $(H, W)$ dimension and $\mathbf{X}_{mff}$ in channel dimension, and use a 1×1 convolution $\mathrm{W}_\Upsilon$ to obtain $\Upsilon$ .

\subsubsection{Viewport-wise Perceptual Quality Prediction Module}
\paragraph{Viewports Vit}
Inspired by the recency effect of the HVS~\cite{Sui,recency}, we design the viewports vit (VV) module to integrate the quality of all viewports to get whole perceived quality of OIs. Specifically, we first flatten $\Upsilon$ in $ (M, \frac{HW}{64}) $ dimension to fuse different information of each viewport and obtain $\mathbf{F} \in \mathbb{R}^{ \frac{MHW}{64} \times \hat{C}}$. Then, the multi-head attention~\cite{vaswani2017attention} is used to extract the correlation information between these features and promote the integration of global information. The whole process can be described as 
\begin{eqnarray}
\centering 
\mathbf{VF}=\mathrm{MultiHead}(\mathbf{F}) = \mathrm{Cat}(\mathbf{Head}_1, \cdot\cdot\cdot, \mathbf{Head}_i)\mathbf{W}_O,
\label{eq:feature}
\end{eqnarray}

\begin{eqnarray}
\centering 
\mathbf{Head}_i= \mathrm{Softmax}\left(\frac{\mathbf{F} \cdot \mathbf{W}_i^Q (\mathbf{F} \cdot \mathbf{W}_i^K)^T}{\sqrt{d_k}}\right)\mathbf{F} \cdot \mathbf{W}_i^V,
\label{eq:feature}
\end{eqnarray}
where $\mathbf{W}_O$ is output weight matrix; \(\mathbf{W}_i^Q, \mathbf{W}_i^K, \mathbf{W}_i^V\) are the weight matrices associated with the \(i\)-th attention head; \(d_k\) is the dimension of $Q$ or $K$; The range of $i$ is from 1 to 4.

Two linear layers and Gaussian error linear unit (GELU) function are used to map $\mathbf{VF} \in \mathbb{R}^{ \frac{MHW}{64} \times \hat{C}}$ into a vector $\mathbf{o} \in \mathbb{R}^{\frac{MHW}{64} \times 1}$, and $\mathbf{o}$ is integrated and compressed by the transposition operation and linear layer. Finally, the perceptual quality score $q$ of the whole OI can be obtained by 
\begin{eqnarray}
\centering 
q=\mathrm{Linear}(\mathrm{Transpose}(\mathrm{MLP}(\mathbf{VF}))),
\label{eq:feature}
\end{eqnarray}

\paragraph{Loss Function}

In this study, we use $\mathrm{L}$2 loss function to train and optimize the model, which is defined as

\begin{eqnarray}
\centering 
\mathrm{L}=\frac{1}{N} \sum_{i=1}^{N} (\hat{q}_i - q_i)^2,
\label{eq:feature}
\end{eqnarray}
where $q_i$ and $\hat{q}_i$ are the predicted quality score and the subjective score of the $i$-th OI, respectively.

\begin{table*}
\caption{Performance comparison of state-of-the-art 2D-IQA and OIQA methods on the proposed database. In each column, the best results are highlighted in bold. * denotes the model adopts the equatorial sampling viewport extraction method as input.}
\begin{tabular}{m{0.8cm}<{\centering} m{2.5cm}<{\centering}
 m{0.45cm}<{\centering} m{0.45cm}<{\centering} m{0.7cm}<{\centering} m{0.45cm}<{\centering} m{0.45cm}<{\centering} m{0.7cm}<{\centering} m{0.45cm}<{\centering} m{0.45cm}<{\centering} m{0.7cm}<{\centering} m{0.45cm}<{\centering} m{0.45cm}<{\centering} m{0.7cm}<{\centering} m{0.45cm}<{\centering} m{0.45cm}<{\centering} m{0.7cm}<{\centering} }
\toprule
\multirow{4}{*}{\centering Type} & \multirow{4}{*}{Metrics} & \multicolumn{3}{c}{BD}  & \multicolumn{3}{c}{GB}   & \multicolumn{3}{c}{GN}   & \multicolumn{3}{c}{ST}   & \multicolumn{3}{c}{Overall} \\ \cmidrule(lr){3-5}\cmidrule(lr){6-8}\cmidrule(lr){9-11}\cmidrule(lr){12-14}\cmidrule(lr){15-17}
                      &   & PLCC & SRCC  & RMSE & PLCC & SRCC & RMSE & PLCC & SRCC & RMSE & PLCC & SRCC & RMSE & PLCC & SRCC & RMSE \\
\midrule
\multirow{8}{*}{\makecell[c]{2D-\\IQA}} & NIQE~\cite{mittal2012making} & 0.226 & 0.199 & 0.602 & 0.147 & 0.147 & 0.596 & 0.113 & 0.047 & 0.517 & 0.139 & 0.090 & 0.554 & 0.094 & 0.047 & 0.605\\[2pt]
                      & HyperIQA~\cite{su2020blindly} & 0.267 & 0.265 & 0.595 & 0.361 & 0.353 & 0.561 & 0.256 & 0.252 & 0.502 & 0.122 & 0.102 & 0.555 & 0.201 & 0.198 & 0.595\\[2pt]
                      & UNIQUE~\cite{zhang2021uncertainty} & 0.209 & 0.204 & 0.604 & 0.285 & 0.277 & 0.577 & 0.380 & 0.374 & 0.481 & 0.123 & 0.109 & 0.555 & 0.174 & 0.169 & 0.599 \\[2pt]
                      & CONTRIQUE~\cite{madhusudana2022image} & 0.168 & 0.157 & 0.609 & 0.036 & 0.029 & 0.602 & 0.050 & 0.051 & 0.519 & 0.052 & 0.016 & 0.559 & 0.089 & 0.076 & 0.606\\[2pt]
                      & MANIQA~\cite{yang2022maniqa} & 0.009 & 0.008 & 0.618 & 0.126 & 0.109 & 0.598 & 0.287  &  0.279  & 0.48 & 0.124 & 0.073 & 0.555 & 0.101 & 0.090 &  0.605\\[2pt]
                      & VCRNet~\cite{pan2022vcrnet} & 0.222 & 0.214 & 0.602 & 0.339 & 0.340 & 0.566 & 0.092  &  0.094  & 0.517 &
                      0.052 & 0.042 & 0.558 & 0.162  &  0.150 &  0.599\\[2pt]
                      & LIQE~\cite{zhang2023blind} & 0.031 & 0.028 &  0.618 & 0.036 & 0.039 & 0.601 & 0.494  &  0.481  & 0.452 & 0.105 & 0.005 & 0.556 & 0.029  &  0.026 &  0.607\\[2pt]
                      & SAAN~\cite{yi2023towards} & 0.184 & 0.170 & 0.607 & 0.095 & 0.088 & 0.599 & 0.001 &  0.013 & 0.519 & 0.087 & 0.087 & 0.557 & 0.056 & 0.040 & 0.607\\[2pt]
\midrule
\multirow{4}{*}{\makecell[c]{FR-\\OIQA}}   & S-PSNR~\cite{s-psnr}    & 0.692 & 0.694 & 0.446  & 0.413 & 0.276 & 0.549  & 0.448 & 0.275 & 0.465  & 0.152 & 0.130 & 0.553 & 0.355 &0.285 & 0.568\\[2pt]
                      & WS-PSNR~\cite{ws-psnr}   & 0.686 & 0.688 & 0.450 & 0.412 & 0.275 & 0.549  & 0.449 & 0.279 & 0.465 & 0.141 & 0.114 &0.554 & 0.353 & 0.284 &0.569\\[2pt]
                      & CPP-PSNR~\cite{cpp-psnr}  & 0.686 & 0.688 & 0.450  & 0.412 & 0.275 & 0.549 & 0.448 & 0.274 & 0.465 & 0.140 & 0.114 & 0.554 & 0.355 & 0.285 & 0.568  \\[2pt]
                      & WS-SSIM~\cite{ws-ssim} & 0.415 &  0.310 & 0.562 & 0.462  &  0.312  &  0.534 & 0.448 & 0.289 &   0.465 & 0.110 &0.055 & 0.556 & 0.388 & 0.249  & 0.560\\[2pt]
\midrule
\multirow{7}{*}{\makecell[c]{NR-\\OIQA}}   & MC360IQA*~\cite{MC360IQA}   & 0.807 & 0.816 & 0.374  & 0.791 & 0.793 & 0.364  & 0.714 & 0.730 & 0.373  & 0.221 & 0.200 & 0.523  & 0.620 & 0.611 & 0.474\\[2pt]
  & MC360IQA~\cite{MC360IQA}   & 0.831 & 0.833 & 0.353 & 0.786 & 0.785 & 0.367 & 0.732 & 0.747 & 0.363 & 0.334 & 0.304 & 0.505 & 0.689 & 0.683 & 0.437\\[2pt]
  & VGCN*~\cite{VGCN}    & 0.844 & 0.845 & 0.341  & 0.572 & 0.568 & 0.488  & 0.515 & 0.512 & 0.458  & 0.274 & 0.265 & 0.515  & 0.464  & 0.367 & 0.535 \\[2pt]
  & ST360IQ~\cite{jabbari2023st360iq}    & 0.750 & 0.758 & 0.427  & 0.707 & 0.708 & 0.426  & 0.707 & 0.717 & 0.384  & 0.336 & 0.331 & 0.552  & 0.640  & 0.632 & 0.481 \\[2pt]
  & Fang22*~\cite{fang2022perceptual}    & 0.846 & 0.846 & 0.338  & 0.812 & 0.813 & 0.347  & 0.786 & 0.798 & 0.330  & 0.401 & 0.374 & 0.491  & 0.633  & 0.616 & 0.468 \\[2pt]
  & Assessor360~\cite{wu2023assessor360}   & 0.722 & 0.720 & 0.439 & 0.681 & 0.691 & 0.435 & 0.704 & 0.717 & 0.379 & 0.533 & 0.532 & 0.453 & 0.694 & 0.690 & 0.434 \\[2pt]
    & OIQAND    & \textbf{0.882} & \textbf{0.885} & \textbf{0.298}  & \textbf{0.831} &\textbf{0.834} & \textbf{0.331}  & \textbf{0.828} & \textbf{0.820} & \textbf{0.314}  & \textbf{0.683} & \textbf{0.681} & \textbf{0.391}  & \textbf{0.800}  & \textbf{0.800} & \textbf{0.362} \\[2pt]
\bottomrule
\end{tabular}

\label{tab-performanceDB}
\end{table*}

\subsection{Implementation Details}
\label{sec:implement}

We split the OIs in the JUFE-10K database into $80\%$ for training and $20\%$ for testing. The size of the input viewport is set to $224\times224\times3$. The number of channels $C1$, $C2$, $C3$, and $C4$ are 256, 512, 1024, and 1024. The proposed model is trained on a high-performance server with Inter(R) Xeon(R) Gold 6326 CPU@2.90GHz, 24G NVIDIA GeForce RTX A5000 GPU, and 260GB RAM. The model is optimized using the Adaptive moment estimation (Adam) optimizer with an initial learning rate of $10^{-5}$. During training, the batch size is set to 24 and the training process ends after 25 epochs.

\section{Experiments}
\label{sec:qu_ex}
In this section, we first introduce the evaluation criteria and then show performance comparison of the proposed OIQAND and state-of-the-art 2D-IQA and OIQA methods on the proposed database. Finally, we carry out ablation studies on several key factors regarding the performance of the proposed model.

\subsection{Evaluation Criteria}
\label{subsec:ex_se}

Three standard performance criteria, including Pearson's linear correlation coefficient (PLCC), Spearman's rank order correlation coefficient (SRCC) and root mean square error (RMSE), are used to measure the prediction monotonicity and accuracy. An OIQA model with better performance has higher values of PLCC and SRCC and lower RMSE value. As suggested in~\cite{vqeg}, the predicted scores are first mapped to subjective ratings before calculating PLCC and RMSE. We adopt the following four-parameter logistic function:
\begin{eqnarray}
\centering 
f(q)=\frac{\eta_1-\eta_2}{1+e^{(-\frac{q-\eta_3}{\eta_4})}}+\eta_2,
\label{eq:nonlinear_mapping}
\end{eqnarray}
where $\left\{\eta_i|i=1,\cdots,4\right\}$ are parameters to be fitted.

\subsection{Performance Comparison}

We compare the proposed OIQAND with eight 2D-IQA models and nine OIQA models. These 2D-IQA models include NIQE~\cite{mittal2012making}, HyperIQA~\cite{su2020blindly}, UNIQUE~\cite{zhang2021uncertainty}, CONTRIQUE~\cite{madhusudana2022image}, MANIQA~\cite{yang2022maniqa}, VCRNet~\cite{pan2022vcrnet}, LIQE~\cite{zhang2023blind}, and SAAN~\cite{yi2023towards}, and the test OIQA models include S-PSNR~\cite{s-psnr}, WS-PSNR~\cite{ws-psnr}, CPP-PSNR~\cite{cpp-psnr}, WS-SSIM~\cite{ws-ssim}, MC360IQA~\cite{MC360IQA}, VGCN~\cite{VGCN}, Fang22~\cite{fang2022perceptual}, ST360IQ~\cite{jabbari2023st360iq}, and Assessor\cite{wu2023assessor360}, all of them have been introduced in Section~\ref{sec:related}. The experimental results are presented in Table~\ref{tab-performanceDB}, where the best performance is highlighted. As shown in Table~\ref{tab-performanceDB}, we can clearly observe that these 2D-IQA methods perform worse than the OIQA methods, which is attributed to that OIs are significantly different from 2D images. Predicting the quality of OIs by directly using 2D-IQA models ignores the particularity of OIs' geometric characteristics. Among these 2D-IQA models, NIQE only uses luminance features to capture distortion, and it is also limited by few simple-content training images. HyperIQA performs slightly better than others, where the reason may be that HyperIQA fuses local distortion features and global semantic features, and thus is able to capture non-uniform distortion to some extent. LIQE shows overall poor performance, which is due to that LIQE randomly crops multiple sub-images from the test OI as input, and may extract the areas without distortion or with global distortion and therefore resulting in opposite quality prediction. SAAN focuses on artistic image aesthetic assessment, which has natural gap with OIQA. CONTRIQUE may be biased to more high-quality regions due to its cropping operation, and therefore affect its performance. As for FR-OIQA methods, \ie, S-PSNR, WS-PSNR, CPP-PSNR, and WS-SSIM, we can clearly observe that FR-OIQA models perform better than 2D-IQA models, which is reasonable since they are designed on PSNR and SSIM by considering the geometry characteristic of OIs. Moreover, these FR-OIQA methods show worse results on the \textit{locally distributed} ST distortion than other \textit{globally distributed} distortions, which demonstrates the necessity of considering viewing behavior in modeling the quality of OIs. As for NR-OIQA methods, including MC360IQA, VGCN, Fang22, ST360IQ, and Assessor360, we retrain them on the JUFE-10K database with the same training and test splitting scheme for a fair comparison. It is worth noting that MC360IQA, VCGN, and Fang22 use our proposed viewport extraction method to get input, while ST360IQ and Assessor360 keep their default settings. Intuitively, MC360IQA performs well except for the ST distortion, which indicates that advanced inter-viewport quality dependence module is necessary for measuring non-uniform distortion. VGCN shows better performance on the BD distortion, while other distortion types perform worse, which may be due to the reduced input of the number of viewports and the fixed viewport area, which leads to the graph convolutional network action not performing well. The overall performance of ST360IQ is slightly better than that of MC360IQA, VGCN, and Fang22, which may benefit from the saliency-guided tangent viewport sampling scheme. From Table~\ref{tab-performanceDB}, we can make several interesting findings. First, although the overall performance of the NR-OIQA method has been significantly improved, there is still much room for improvement. This means that existing models designed to handle uniform distortions cannot effectively handle non-uniform distortions. Second, most of the NR-OIQA methods fail on the ST distortion compared with BD, GB, and GN distortions, which indicates that it is still a challenge to deal with the unique ST distortion of OIs. Third, Transformer-based models outperform other models in general. This indicates that the transformer is better at capturing local distortion.

The proposed OIQAND model outperforms these state-of-the-art models on the proposed JUFE-10K database. Specifically, OIQAND outperforms Assessor360 by about 10\% in overall performance. The performance of OIQAND on the BD, GB, GN, and ST distortions is better than that of Fang22 and Assessor360 models by 3.6\%, 1.9\%, 4.2\%, and 15\%. Fang22 achieves significant improvement in the BD, GB, and GN distortions by introducing users' viewing conditions. Assessor360 achieves better overall performance by introducing gated recurrent unit (GRU) modules to model the relationship of viewport sequences. In general, similar to other models, the proposed OIQAND model shows less satisfactory performance when facing with the ST distortion. One possible reason is that the adopted viewport extraction method undersamples the locally distorted region(s), and samples much more viewports from high-quality regions. Therefore, the advanced human behavior simulating model is desired to be deeply explored, and can be embedded into OIQAND for further improvement. Furthermore, although the proposed DAP and MFF modules present high effectiveness in capturing non-uniform distortion (which will be discussed in Section~\ref{subsec:as}), more efforts are expected to be put into designing dynamic fusion module for better capturing non-uniform distortion.

\subsection{Ablation Studies}
\label{subsec:as}
In this subsection, we conduct a series of ablation experiments on the JUEF-10K database to explore the effectiveness of those important modules.

\paragraph{Effectiveness of each module} 

\begin{table}[H]
\caption{Ablation study results of each module in the OIQAND model. Backbone: swin transformer.}
\centering
\setlength{\tabcolsep}{4pt}
\begin{tabular}{cccc|ccc}
\hline
\toprule
Backbone & MFF & DAP & VPQP & PLCC & SRCC & RMSE \\
\midrule
\checkmark &     &     &       & 0.421 & 0.419 & 0.546 \\
\checkmark  & \checkmark   &     &       & 0.431 & 0.409 & 0.544 \\
\checkmark  &     & \checkmark  &       & 0.788 & 0.785 & 0.371 \\
\checkmark  &     &     & \checkmark     & 0.778 & 0.775 & 0.378 \\
\checkmark  & \checkmark  & \checkmark   &       & 0.789 & 0.789 & 0.370 \\
\checkmark  & \checkmark  & \checkmark  & \checkmark     & \textbf{0.800} & \textbf{0.800} & \textbf{0.362} \\
\bottomrule
\end{tabular}
\label{tab:tab-multiscale}
\end{table}

The results are shown in Table~\ref{tab:tab-multiscale}, where we can find that both the DAP module and the VPQP module contribute significantly to the performance. Integrating the MFF, DAP, and VPQP modules can further improve the performance. This manifests that the combination of multi-scale features of viewports and adaptive content distortion perception and multi-viewport fusion can successfully capture non-uniform distortions and effectively predict the perceptual quality of OIs.

\begin{table}[H]
\caption{Effectiveness of the non-uniform distortion perception module in the OIQAND model. w/o: without. w/: with.}
\centering
\begin{tabular}{c|ccc}
\hline
\toprule
Method              & PLCC  & SRCC  & RMSE   \\
\midrule
w/o VAC             & 0.787 & 0.786 & 0.372 \\
w/o ACAC            & 0.783 & 0.782 & 0.375 \\
w/o VV              & 0.789 & 0.789 & 0.370 \\
w/o VAC + ACAC      & 0.780 & 0.779 & 0.377 \\
w/o VAC + ACAC + VV & 0.430 & 0.421 & 0.540 \\
w/ VAC + ACAC + VV  & \textbf{0.806} & \textbf{0.804} & \textbf{0.357} \\
\bottomrule
\end{tabular}
\label{tab:tab-NUdistortion}
\end{table}

\paragraph{Effectiveness of non-uniform distortion sensing modules} 
To investigate the effectiveness of the modules designed specifically for non-uniform distortion, we conduct the ablation study, and the experimental results are shown in Table~\ref{tab:tab-NUdistortion}. We can observe that when we remove one or both of the VAC, ACAC, and VV modules, the performance is not drastically degraded. When we remove these three modules VAC, ACAC, and VV together, the performance shows a significant degradation. This indicates that the performance of OIQAND largely depends on the cooperation of these non-uniform distortion sensing modules.

\begin{table}[H]
\caption{Results of using different viewport generation as input in the OIQAND model.}
\centering
\begin{tabular}{c|c|ccc}
\hline
\toprule
Generate viewports  & Num & PLCC & SRCC & RMSE \\
\midrule
Spherical sampling~\cite{fang2022perceptual}  & 20  &  \textbf{0.821}    &  \textbf{0.821}    &    \textbf{0.344}  \\
Spherical sampling~\cite{fang2022perceptual}  & 8  &  0.813   &  0.810   &   0.352 \\
Saliency~\cite{VGCN}            & 20  &   0.820   &   0.820   &   0.345   \\ 
Saliency~\cite{VGCN}            & 8  &   0.775   &   0.776   &   0.382   \\ 
ScanDMM~\cite{Sui}             & 15  &   0.752   &   0.751   &  0.397  \\ 
RPS~\cite{wu2023assessor360}                 & 8   &   0.757   &   0.752   &   0.394   \\ 
Equatorial sampling & 20   &   0.813  &   0.812   &   0.352  \\
Equatorial sampling & 8   &   0.800  &   0.800   &   0.362  \\
\bottomrule
\end{tabular}
\label{tab:tab-viewport}
\end{table}

\paragraph{Influence of viewport generation methods} 
Different viewport content and the number of inputs have different effects on the performance of the proposed OIQAND. We explore the sensitivity of OIQAND to different viewport generation methods. As shown in Table~\ref{tab:tab-viewport}, we can observe that spherical sampling~\cite{fang2022perceptual} and saliency~\cite{VGCN} with 20 viewports perform better, probably because as the number of viewports increases, the model can capture more view content and further understand the perception of the whole image. Note that the amount of computation increases as the number of viewports increases. Although the number of viewports increases, the performance of scanDMM~\cite{Sui} decreases, which may be due to excessive information redundancy in the viewport content extracted by scanpath. RPS~\cite{wu2023assessor360} has the same number of viewports as the equatorial sampling in this paper, but its performance is lower, which may be due to the incomplete range of viewport content extracted by RPS~\cite{wu2023assessor360}, and leads to a poor understanding of the overall image distortion effect by the model. Overall, the adopted equatorial sampling strategy shows good competitiveness when the number of viewport images is set to 8 or 20.

\section{Conclusion}
\label{sec:conc}

In this paper, we comprehensively study the perceptual quality assessment of non-uniformly distorted OIs from both subjective and objective perspectives. Specifically, we construct a novel non-uniformly distorted OIQ database,~\ie, JUFE-10K, which has 10,320 non-uniformly distorted OIs, and explore the effect of viewing conditions on the perceptual quality of OIs without setting fixed viewing starting point and exploration time. From subjective experiment, we find that the viewing initial point has little impact on the perceived quality of OIs. Furthermore, we propose a perception-guided OIQA model named OIQAND for non-uniform distortion, which uses the swin transformer as backbone, two parallel modules,~\ie, a distortion adaptive perception module and a multi-scale feature fusion module, for capturing viewport-wise distortion jointly, and a viewport-wise perceptual quality prediction model for mapping the distortion-aware features of viewports to global quality score. Extensive experiments on the proposed JUFE-10K database show that the proposed model shows excellent performance compared with state-of-the-art 2D-IQA and OIQA methods, and demonstrate the effectiveness of each module. In the future, we will consider the full utilization of the database, \eg, investigating authentic scanpath prediction for OIs based on the collected EM and HM data, or exploring scanpath prediction jointly with quality assessment.

%



\ifCLASSOPTIONcaptionsoff
  \newpage
\fi



%

\bibliographystyle{IEEEtran}
\bibliography{reference}

\vspace{-10 mm} 
\begin{IEEEbiography}[{\includegraphics[width=1.0in,height=1.3in,clip]{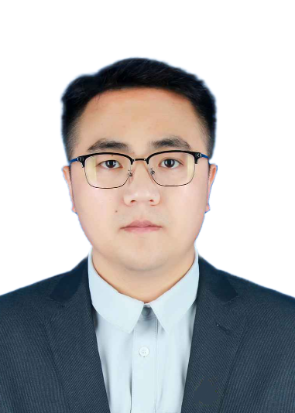}}]{Jiebin Yan} received the Ph.D. degree from Jiangxi University of Finance and Economics, Nanchang, China. He was a computer vision engineer with MTlab, Meitu. Inc, and a research intern with MOKU Laboratory, Alibaba Group. From 2021 to 2022, he was a visiting Ph.D. student with the Department of Electrical and Computer Engineering, University of Waterloo, Canada. He is currently a Lecturer with the School of Computing and Artificial Intelligence, Jiangxi University of Finance and Economics, Nanchang, China. His research interests include visual quality assessment and computer vision.
\end{IEEEbiography}
\vspace{-10 mm} 

\begin{IEEEbiography}[{\includegraphics[width=1.0in,height=1.3in,clip]{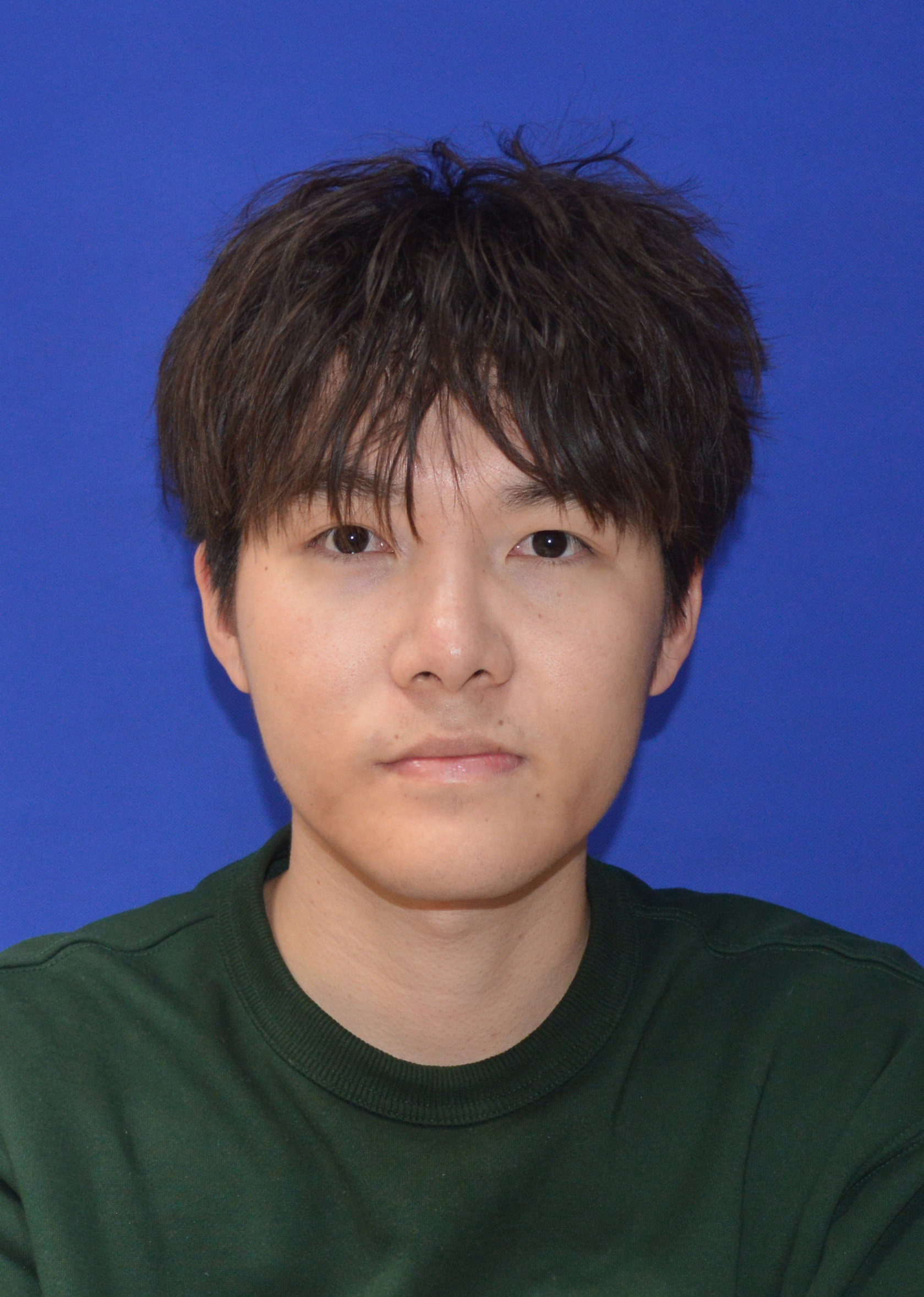}}]{Jiale Rao} received the B.E. degree from the Jiangxi Agricultural University, Nanchang, China, in 2022,  He is currently working toward the M.S. degree with the School of Computing and Artificial Intelligence, Jiangxi University of Finance and Economics. His research interests include visual quality assessment and VR image processing.
\end{IEEEbiography}
\vspace{-10 mm} 

\begin{IEEEbiography}[{\includegraphics[width=1.0in,height=1.3in,clip]{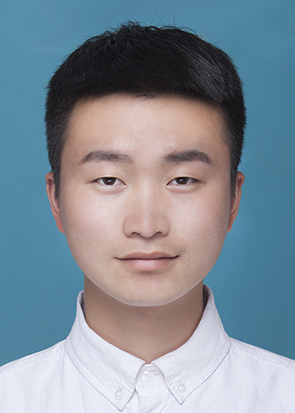}}]{Xuelin Liu} received the B.E., M.S. and Ph.D. degrees from the Jiangxi University of Finance and Economics, Nanchang, China, in 2017, 2020 and 2024, respectively. He is currently a post-doctoral fellow at Jiangxi University of Finance and Economics. From May 2023 to June 2024, he was a visiting PhD student at the City University of Hong Kong. His research interests include visual quality assessment and VR image/video processing.
\end{IEEEbiography}
\vspace{-10 mm} 

\begin{IEEEbiography}[{\includegraphics[width=1.0in,height=1.3in,clip]{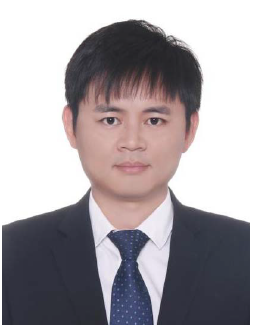}}]{Yuming Fang}(S’13–SM’17) received the B.E. degree from Sichuan University, Chengdu, China, the M.S. degree from the Beijing University of Technology, Beijing, China, and the Ph.D. degree from Nanyang Technological University, Singapore. He is currently a Professor with the School of Computing and Artificial Intelligence, Jiangxi University of Finance and Economics, Nanchang, China. His research interests include visual attention modeling, visual quality assessment, computer vision, and 3D image/video processing. He serves on the editorial board for \textsc{IEEE Transactions on Multimedia} and \textsc{Signal Processing: Image Communication}.
\end{IEEEbiography}

\vspace{-10 mm} 
\begin{IEEEbiography}[{\includegraphics[width=1.0in,height=1.3in,clip]{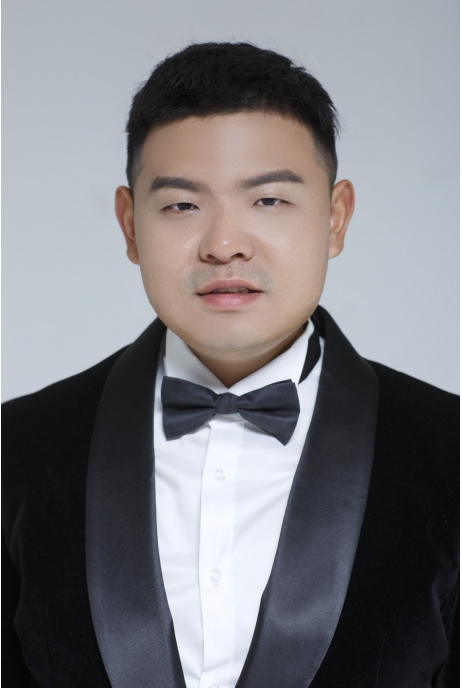}}]{Yifan Zuo} (Member, IEEE) received the Ph.D. degree from the University of Technology Sydney, Ultimo, NSW, Australia, in 2018. He is currently an Associate Professor with the School of Computing and Artificial Intelligence, Jiangxi University of Finance and Economics. His research interests include computer vision, image processing. The corresponding papers have been published in important international conferences including ICIP, ICME, and major international journals such as IEEE TRANSACTIONS ON IMAGE PROCESSING, IEEE TRANSACTIONS ON CIRCUITS AND SYSTEMS FOR VIDEO TECHNOLOGY, and IEEE TRANSACTIONS ON MULTIMEDIA.
\end{IEEEbiography}

\vspace{-10 mm} 
\begin{IEEEbiography}[{\includegraphics[width=1.0in,height=1.3in,clip]{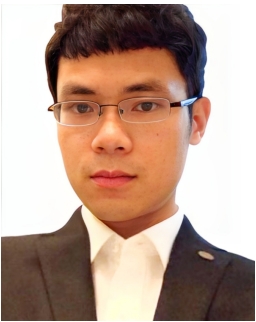}}]{Weida Liu} is currently a Research Fellow at Harvard Medical School, Harvard University. Before that, he was a Research Scientist at A*STAR, Singapore. Weide received his Ph.D. from Nanyang Technological University. His research interests include computer vision, language, machine learning, federated learning, and medical image analysis.
\end{IEEEbiography}

\end{document}